\documentclass[lettersize,journal]{IEEEtran}

\usepackage{amsfonts,amssymb,amsbsy,amsmath,paralist,theorem,bm,ifthen,color}
\usepackage{graphicx}
\usepackage{subfigure,relsize}

\usepackage[noadjust]{cite}
\usepackage[numbers,sort&compress,square]{natbib}

\usepackage{cases,booktabs}
\usepackage{flushend}

\usepackage{algorithmic,algorithm}

\usepackage{lipsum}

\newtheorem{Theorem}{Theorem}

\newtheorem{Proof}{Proof}

\newtheorem{Rmk}{Remark}

\newcommand\Sc{\ensuremath{\mathcal{S}}}

\newcommand\Lc{\ensuremath{{\mathcal{L}}}}

\newcommand\Ac{\ensuremath{{\mathcal{A}}}}

\newcommand\Ec{\ensuremath{{\mathcal{E}}}}
\newcommand\Kc{\ensuremath{{\mathcal{K}}}}

\newcommand\Uc{\ensuremath{{\mathcal{U}}}}

\newcommand\Ebb{\ensuremath{{\mathbb{E}}}}

\newcommand\wb{\ensuremath{{\bm w}}}
\newcommand\yb{\ensuremath{{\bm y}}}

\newcommand\sbb{\ensuremath{{\bm s}}}

\newcommand\hb{\ensuremath{{\bm h}}}

\newcommand\Ab{\ensuremath{{\bm A}}}
\newcommand\ab{\ensuremath{{\bm a}}}

\newcommand\bb{\ensuremath{{\bm b}}}
\newcommand\Cb{\ensuremath{{\bm C}}}

\newcommand\Db{\ensuremath{{\bm D}}}

\newcommand\fb{\ensuremath{{\bm f}}}

\newcommand\Ib{\ensuremath{{\bm I}}}
\newcommand\Ic{\ensuremath{{\mathcal{I}}}}

\newcommand\rb{\ensuremath{{\bm r}}}

\newcommand\vb{\ensuremath{{\bm v}}}
\newcommand\Wb{\ensuremath{{\bm W}}}

\newcommand\zb{\ensuremath{{\bm z}}}
\newcommand\alphab{\ensuremath{{\bm \alpha}}}

\newcommand\E{\ensuremath{{\mathbb{E}}}}

\newcommand\Cbb{\ensuremath{{\mathbb{C}}}}
\newcommand\Rbb{\ensuremath{{\mathbb{R}}}}

\DeclareMathOperator*{\argmax}{argmax}

\graphicspath{{figures/}}

\definecolor{applegreen}{rgb}{0.55, 0.71, 0.0}

\begin{document}
	
	\title{Toward Distributed User Scheduling and Coordinated Beamforming in Multi-Cell mmWave Networks: A Sensing-Assisted Framework  
	}
	
	\author{Tenghao~Cai, Lei~Li,  \textit{Member, IEEE}, Shutao~Zhang and Tsung-Hui~Chang,  \textit{Fellow, IEEE} \vspace{-0.6cm}
		
		\thanks{\smaller[1] T. Cai is with the School of Science and Engineering, The Chinese University of Hong Kong, Shenzhen (CUHK-Shenzhen), China, and with the Shenzhen Research Institute of Big Data (SRIBD). L. Li and T.-H. Chang are with School of Artificial Intelligence, CUHK-Shenzhen and with the SRIBD. S. Zhang is with the Networking and User Experience Laboratory, Huawei Technologies, Shenzhen, China (email: 221019048@link.cuhk.edu.cn, lei.ap@outlook.com, zhangshutao2@huawei.com, tsunghui.chang@ieee.org). An earlier version of this paper was presented in part at the IEEE ICASSP 2024 \cite{SD_USBF_icassp24}.}

	}

	\maketitle

	\begin{abstract}
		Providing guaranteed quality of service for cell-edge users remains a longstanding challenge in wireless networks. While coordinated interference management was proposed decades ago, its potential has been limited by computational complexity and backhaul resource constraints. Distributed user scheduling and coordinated beamforming (D-USCB) offers a scalable solution but faces practical challenges in acquiring inter-cell channel state information (CSI), as base stations (BSs) are often restricted to signal strength measurements, and high-dimensional CSI exchange incurs substantial overhead. Inspired by integrated sensing and communication (ISAC), this paper proposes a sensing-assisted D-USCB (SD-USCB) framework to maximize the network throughput of multi-cell mmWave networks. Firstly, the framework leverages channel knowledge maps (CKMs) that map user locations to CSI estimates, where user locations are proactively sensed via ISAC echoes. Secondly, we employ a signal-to-average-leakage-plus-interference-plus-noise ratio (SALINR) metric for distributed ISAC beamforming optimization, in which BSs simultaneously communicate with users and sense their locations. These two components jointly enable distributed coordinated transmission with only user location information exchanged among BSs, thereby substantially reducing backhaul overhead. In addition, we devise efficient distributed user scheduling and ISAC beamforming algorithms to jointly optimize communication and sensing performance. Extensive numerical results demonstrate significant improvements in network throughput, validating the efficacy of the proposed framework. 
		
		\vspace{0.3cm}
	\end{abstract}

	\IEEEpeerreviewmaketitle

	\vspace{-0.2cm}
	\section{Introduction} \label{sec:intro}
	\vspace{-0.0cm}

	{

		Multi-cell coordinated resource allocation (MCRA) has been widely recognized as a promising approach to improving the cell-edge performance of wireless networks \cite{li2015multicell}. By judiciously designing user scheduling and transmit beamforming (BF), MCRA can effectively mitigate inter-user interference and exploit the spatial degrees of freedom to enhance the capacity in multi-cell networks. Traditional centralized MCRA (C-MCRA) schemes typically require a central unit (CU) to collect global channel state information (CSI) from base stations (BSs) via backhauls to perform global optimization \cite{he2022joint}. However, with the increased number of antennas, the channel dimension expands, and the CSI collection in C-MCRA can introduce significant overhead, given that both the intra-cell CSI of each BS and the inter-cell CSI are required. Moreover, the computational complexity of C-MCRA schemes escalates with the growing number of BSs and users, making them poorly scalable to larger networks \cite{xu2024distributed}. 
		
		In contrast, leveraging iterative distributed optimization methods such as primal/dual decomposition, alternating direction method of multipliers (ADMM), or the virtual interference approach, distributed MCRA (D-MCRA) schemes \cite{liu2024survey} were proposed to enable distributed optimization and parallel computation. By introducing local variable copies and consensus variables, the original centralized problem can be decomposed into smaller local subproblems, each solved at individual BSs. These subproblems are then coordinated to obtain a solution to the original large problem through iterative updates of local and consensus variables. As the optimization is distributed, the computational burden at each BS is effectively reduced. However, these methods are still not satisfactory in practical systems as they require a large number of iterations and information exchange across BSs.

		In addition, these D-MCRA methods face several practical challenges, including {\bf i) CSI acquisition:} most existing D-MCRA methods assume that both intra-cell CSI and inter-cell CSI are available beforehand, whereas acquiring such information is highly non-trivial in practical systems. Specifically, the intra-cell CSI can be obtained via pilot-based estimation at the user, which then feeds the estimate back to the BS. The inter-cell CSI, however, is not always available at the BS. According to \cite{TS_38214, TS_36214}, in the coordinated transmission mode, the neighboring BS can only obtain the reference signal received power (RSRP) measurements of the inter-cell channels through user feedback\footnote{For a user with joint transmission from multiple BSs, it will feed  CSI back to multiple BSs, but not in the case of coordinated transmission.}.  Consequently, BSs only have access to the signal strength rather than full inter-cell CSI estimates.
		{\bf ii)  substantial overhead of information exchange:} including inter-cell CSI exchange before optimization and intermediate variable (e.g., consensus variable) updates during iterations. In practice, given the limited bandwidth of backhauls (e.g., IP-RAN) connecting BSs, frequent exchange of high-dimensional information imposes significant overhead and delays, leading to outdated information and degraded network performance.
		
		\begin{figure}[t] 
			\centering	
			{\includegraphics[width=0.48\textwidth]{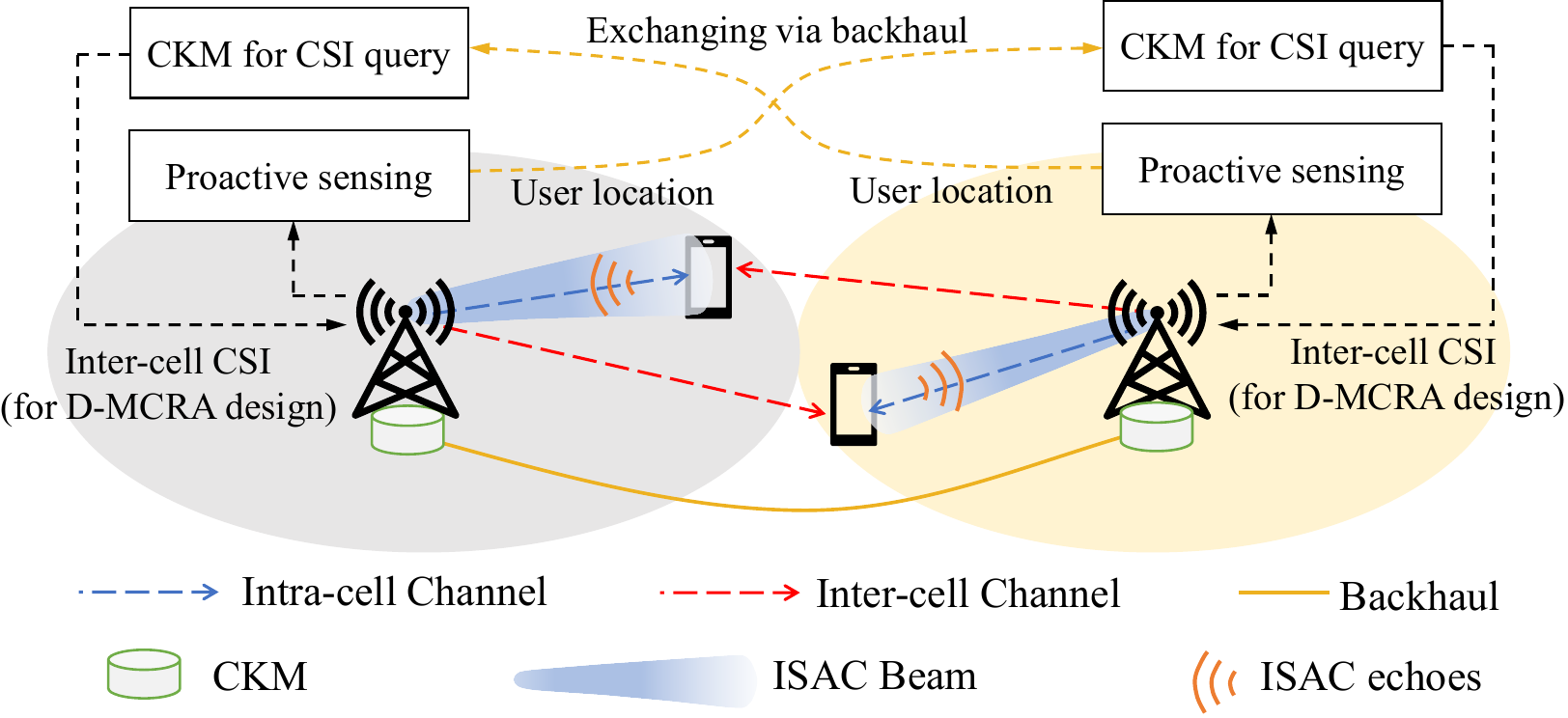}}\vspace{-0.15cm}	
			\caption{
				{\bf Proposed SD-USCB framework:} BSs in the network transmit ISAC signals to simultaneously convey the data to users and sense the user locations.  The sensed locations are then exchanged through the backhaul to enable neighboring BSs to know the locations of active users. With the exchanged locations, BSs query the pre-built CKMs to obtain the inter-cell CSI estimates for subsequent D-MCRA designs.}
			\label{fig:D_USCB} \vspace{-0.05cm}
		\end{figure}
		
		In this work, we investigate the D-MCRA design in multi-cell mmWave networks, aiming to maximize the network throughput characterized by the proportional fairness rate (PFR) via distributed user scheduling and coordinated BF (D-USCB) with significantly reduced information exchange overhead. To address the 1st challenge of inter-cell CSI acquisition, we propose a \textit{sensing-assisted distributed USCB (SD-USCB)} framework. As shown in Fig. \ref{fig:D_USCB}, in this framework, a channel knowledge map (CKM) \cite{CKM_tut24_Zengyong}, which is a BS-specific database mapping user locations to their corresponding (statistical) channel knowledge, is built from the RSRP measurements at each BS to provide the inter-cell CSI statistics like the channel angular power spectrum (APS) \cite{zhang2023physics}. Integrated sensing and communication (ISAC) transmission is then introduced to enable BSs to proactively sense the users' locations, which are input to the CKM for the inter-cell CSI query.  
		
    		To cope with the 2nd challenge of substantial inter-cell information exchange, we introduce a leakage-based metric named `signal-to-average-leakage-plus-interference-plus-noise ratio (SALINR)'  \cite{cai2023approaching} as a surrogate of signal-to-interference-and-noise-ratio (SINR). This transforms the required inter-cell CSI  from \textit{incoming} inter-cell channels to \textit{outgoing} inter-cell channels\footnote{The \textit{incoming} channels to BS-A refer to the inter-cell interference (ICI) channels from BS-B to the users served by BS-A, which, from the perspective of BS-B, are termed \textit{outgoing} inter-cell channels.}. With available outgoing CSI, SALINR-based approximation enables fully distributed optimization, eliminating the cross-cell exchange of intermediate variables during optimization iterations required by existing distributed methods \cite{liu2024survey, padmanabhan2019distributed}. Most importantly, by incorporating the SALINR-based optimization into the SD-USCB framework, each BS only shares its scheduled users’ locations with the other BSs, enabling them to attain the outgoing inter-cell CSI estimates readily via the CKM. As there is no need to exchange the high-dimensional channel estimates, this significantly reduces the cross-cell backhaul overhead.   
		
		In this framework, we further study user scheduling and coordinated BF designs.  
		First, each BS greedily selects a set of scheduling users using a virtual zero-forcing (ZF) beamformer, which targets users with distinct geographical locations and weak channel correlation.  Secondly, since the coordinated BF impacts both the communication performance and the sensing for user localization, it is formulated as an ISAC SALINR BF optimization problem for PFR maximization subject to sensing error constraints. By employing fractional programming (FP) \cite{shen2019optimization} and successive convex approximation (SCA)  techniques \cite{6363727}, we cope with the non-convexity of the problem and develop an efficient algorithm based on dual optimization. Finally, the CKM construction from the RSRP measurements by leveraging local statistical channel modeling (LSCM) techniques \cite{zhang2023physics} is illustrated.
		Together, our proposed framework effectively improves the network performance while greatly reducing the coordination overhead, offering a promising D-USCB paradigm for multi-cell mmWave networks. 
	}
	
	\vspace{-0.2cm}
	\subsection{Related Works }
	{

		\subsubsection{Multi-cell Coordinated Transmission}
		
		While multi-cell coordinated transmission through joint scheduling and BF can mitigate interference and enhance network performance \cite{yu2013multicell}, traditional C-MCRA approaches require a CU to collect CSI from all BSs \cite{shi2011iteratively},  incurring significant signaling overhead in the cross-cell backhauls and substantial computational burden at the CU, limiting scalability in large networks.
		
		In contrast, D-MCRA decomposes the original problem into local subproblems optimized at individual BSs, reducing complexity and improving scalability. \cite{joshi2014distributed,padmanabhan2019distributed,cai2023approaching}. Leveraging the ADMM algorithm, \cite{joshi2014distributed} replaced the ICI power coupled among BSs with intermediate variables and introduced local variable copies as consensus variables, dividing the power minimization problem into multiple local subproblems. During optimization, each BS solves its local subproblem and exchanges updated variable values iteratively to achieve global consensus. Similarly, the distributed BF algorithm \cite{padmanabhan2019distributed} approximated the performance of centralized schemes by exchanging a `scalarized ICI'. However, these distributed schemes typically require a large number of iterations to converge and iterative inter-BS exchange of variable values, leading to significant signaling overhead and latency, particularly over bandwidth-constrained backhauls.
		
		To address these issues, \cite{cai2023approaching} introduced a distributed BF scheme that only exchanges virtual interference channels and BF powers among BSs in optimization, achieving near-centralized performance with only two rounds of information exchange. 
        %Alternatively, surrogate metrics like signal-to-leakage-and-noise ratio (SLNR) \cite{castaneda2015distributed} and signal-to-leakage-plus-interference-plus-noise ratio (SLINR) \cite{li2022decentralized,han2020distributed} have been proposed to reduce the information exchange across BSs. 
        As a surrogate of SINR, signal-to-leakage-and-noise ratio (SLNR) was introduced \cite{castaneda2015distributed} to manage the interference caused by the desired signal affecting all other users in other cells without considering ICI, eliminating the iterative inter-cell exchange of the intermediate variables. However, since the BF power is scaled equally in both the numerator and denominator, SLNR is less effective for guiding power allocation. Signal-to-leakage-plus-interference-plus-noise ratio (SLINR) extended this by accounting for both intra-cell interference and leakage to users in neighboring cells \cite{li2022decentralized,han2020distributed}. However, SLINR's reliance on coupled scheduling variables makes cross-cell leakage estimation challenging. 
        %The work \cite{li2022decentralized}  addressed this with a traffic model to approximate leakage without requiring other BSs' scheduling decisions.  However, this model can not provide leakage estimates with real-time user scheduling.
        Moreover, discrepancies between leakage-based surrogates (SLNR and SLINR) and SINR could lead to degraded performance for multi-cell transmission.
		
		Critically, all these works \cite{joshi2014distributed,padmanabhan2019distributed,cai2023approaching,castaneda2015distributed,li2022decentralized,han2020distributed} assume full intra-cell and inter-cell CSI availability -- a requirement seldom met in practical 5G frequency division duplex systems, where users can estimate intra-cell channels and return them to their serving BS, but they only measure and report the RSRP for inter-cell channels to neighboring BSs. 
        This gap, combined with the substantial overhead of exchanging high-dimensional inter-cell CSI, presents significant unresolved challenges for real-world D-MCRA implementation.
        
        % Consequently, BSs obtain signal strength measurements rather than full inter-cell CSI, posing a significant challenge for inter-cell CSI acquisition—a topic seldom addressed in existing MCRA research. Further, even with available inter-cell CSI, exchanging it across BSs for D-MCRA optimization incurs substantial backhaul overhead, particularly with the deployment of large antenna arrays. These practical challenges are critical for realizing D-MCRA in real-world systems and need to be considered in relevant designs.
		
		\subsubsection{ISAC and sensing-assisted communication}
		With the potential to bring improved dual-functional performance at reduced cost, ISAC has been recognized as one of the key enabling technologies for next-generation (next-G) wireless networks \cite{dong2022sensing}. 
		The scenarios of most existing ISAC transmission designs can be divided into two main categories: scenarios where sensing targets and communication users are separate and independent \cite{liu2021cramer,zhu2023information,cao2023joint, Kexin_TVT24, Babu_TWC24}, and sensing-assisted communication where users also serve as sensing targets \cite{LiL_WCL22,LiuF_TWC20,Lei_JSAC25}. The work \cite{liu2021cramer} optimized the ISAC BF for a network including a single sensing target and multiple communication users by minimizing the sensing Cramer-Rao bound (CRB) subject to per-user communication SINR constraints. Considering a scenario of bi-static multi-target sensing and multi-user DL communication \cite{zhu2023information}, an ISAC BF algorithm was proposed to minimize the weighted sum of sensing CRB while ensuring the network throughput.  In \cite{Kexin_TVT24}, the joint fronthaul compression and power allocation design was investigated in a networked ISAC system based on the cloud radio access (C-RAN) architecture, where multiple BSs cooperatively localize one target while transmitting data to multiple users. In \cite{Babu_TWC24}, the ISAC precoding for a multi-cell ISAC system with multiple users and a single target per cell is studied, and two algorithms were devised using semidefinite relaxation (SDR) \cite{TomLuo_SDR10} and alternating optimization. However, these ISAC transmission designs \cite{zhu2023information, cao2023joint, Kexin_TVT24, Babu_TWC24} consider separate target(s) and users, assuming the availability of CSI without explaining how it is acquired. 
		
		The 2nd category --  sensing-assisted communication design -- has gained increasing attention. By sensing the propagation environment and/or the kinematic states of users from ISAC echoes, the BS can reduce pilot transmissions and obtain channel estimates more efficiently, especially in LoS-dominant mmWave scenarios. For example, \cite{LiL_WCL22} leveraged ISAC transmissions to sense scatterers, greatly reducing user feedback for accurate CSI recovery. The work \cite{LiuF_TWC20} proposed using ISAC echoes to track vehicular users' kinematic states. Assuming LoS channels and negligible inter-user interference, an EKF-based state tracking scheme was developed for predictive BF, reducing beam tracking overhead effectively. Building on this,  \cite{Lei_JSAC25} incorporated CSI acquisition errors and inter-user interference, formulating a robust ISAC BF problem that minimizes maximum sensing errors while ensuring worst-case communication rates, and developed a series of efficient first-order algorithms in complex propagation environments. While these works \cite{LiL_WCL22,LiuF_TWC20,Lei_JSAC25} offered valuable insights into sensing-assisted communication design, they mainly focused on single-cell scenarios. The research on sensing-assisted communication in multi-cell coordinated ISAC is still in its infancy, where increased interference and coordination complexity demand new frameworks and distributed algorithms for D-MCRA. 
		
		\subsubsection{CKM}
		As mentioned above, efficient CSI acquisition is crucial for adaptive transmission design but remains challenging in practice. With the denser wireless nodes and the advance of more diversified localization techniques, next-G wireless networks are envisioned to have copious sources of location-specific channel data, such as the received signal strength, gains, and angles of channel paths \cite{CKM_WC_21}. As spatial samples of the channel, the data reflects the wireless radio environment \cite{zhang2023physics}. With denser nodes in the network and the fact that a multitude of users would appear repeatedly at nearby locations like roads and stadiums, the granularity of these samples will get finer.  Leveraging the large amount of data, the network can apply powerful data mining and signal processing techniques to extract key wireless propagation features.  Inspired by this, the concept of channel knowledge map was recently proposed \cite{CKM_WC_21, CKM_tut24_Zengyong}, which is a site/BS-specific database with locations of transceivers to provide location-specific channel knowledge.  	
		In \cite{TVT24_BIM_zengyong}, a specific type of CKM -- beam index map -- was constructed to facilitate the mmWave beam alignment with reduced overhead.
		The work \cite{CKM_clutter_ICCC24} proposed a clutter suppression approach to improve the sensing accuracy of ISAC systems by constructing a clutter angle map (CLAM) containing the main clutter angles at each location of interest as a CKM. By extracting the intrinsic wireless channel characteristics from collected samples, CKM provides a promising way for efficient low-overhead CSI acquisition, facilitating environment-aware transmission designs.
		
	}
    \vspace{-0.3cm}
	\subsection{Contributions}
    \vspace{-0.1cm}
	In this work, integrating CKM and ISAC, we propose a novel SD-USCB framework and develop efficient D-USCB algorithms. Our main contributions are summarized as follows:
	
	\begin{enumerate}
		\item To overcome the CSI acquisition and inter-cell information exchange bottlenecks in distributed multi-cell coordinated transmission, we propose a novel SD-USCB framework for sum PFR maximization. This framework leverages ISAC transmission to proactively sense user locations, which are exchanged across BSs and input into constructed CKMs to obtain cross-cell CSI estimates. Compared with exchanging full-dimension CSI, this design greatly reduces cross-cell information exchange. Moreover, by approximating the SINR with our previously proposed leakage-based metric SALINR \cite{cai2023approaching}, the multi-cell coordinated BF problem can be solved fully distributively at each BS, avoiding iterative variable exchange in traditional methods.
		\item We prove that the SALINR-based approximate rate achieves a smaller expected error relative to the true rate than the conventional SLINR-based approximation in Rayleigh channels. This result theoretically confirms, for the first time, the superiority of SALINR over SLINR in SINR approximation.
		
		\item To solve the intricate SALINR-based approximate problem with strong coupling between scheduling and BF variables in SD-USCB, we propose a two-stage scheme. First, an enhanced proportional fairness zero-forcing greedy (PFZFG) algorithm schedules users based on local information in each cell. Then, an ISAC BF optimization problem is formulated to maximize the approximate sum PFR under sensing constraints. To solve it, we apply FP techniques to handle the non-convex objective and SCA techniques to manage non-convex constraints, yielding a convex approximate problem. Exploiting the problem structure, we customize a computationally efficient ISAC BF algorithm based on dual optimization (DualOpt),  which only takes first-order computations and greatly reduces the processing time.
		
		\item By extensive simulations, we demonstrate that our proposed SD-USCB framework significantly outperforms the per-cell scheme while greatly reducing overhead for CSI acquisition and cross-cell information exchange.  Moreover, the SALINR-based approximation delivers distinct performance gains over the conventional SLINR-based approach. Besides, our proposed DualOpt-based BF algorithm achieves comparable performance to the SDR-based counterpart while reducing the computational time by approximately one order of magnitude for a large number of users. 
	\end{enumerate}

	The remainder of the paper is structured as follows. Section \ref{sec:model} introduces the system model and problem formulation. In Section \ref{sec:proposed}, the proposed SD-USCB framework and its corresponding algorithm design are illustrated. The construction of CKM from RSRP is detailed in Section \ref{sec:CKM_cons}. Section \ref{sec:simu} evaluates the proposed designs through extensive numerical simulations. Finally, Section \ref{sec:conclusion} draws the conclusions.

	\vspace{-0.3cm}
	\section{System Model and Problem Description} \label{sec:model} 
	
	\subsection{System Model}
	
	\begin{figure}[t] 
		\centering	
		{\includegraphics[width=0.5\textwidth]{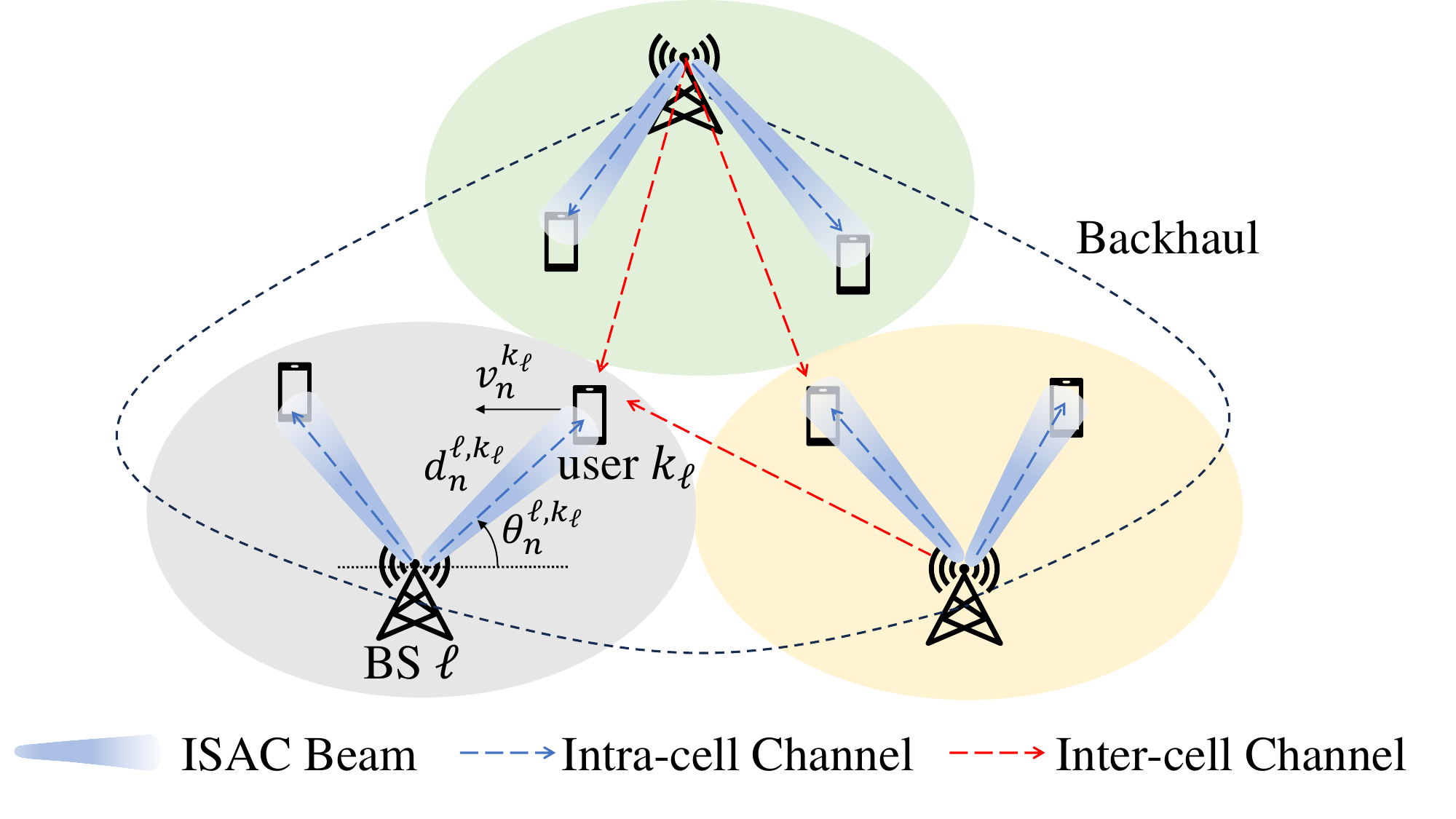}}\vspace{-0.5cm}	
		\caption{A cellular mmWave network, where BSs are inter-connected via limited-bandwidth backhauls. The kinematic states of user $k_\ell \in \Uc_\ell$ include its distance $d_{n-1}^{\ell, k_\ell}$,  angle $\psi_{n-1}^{\ell, k_\ell}$ and  velocity $v_{n-1}^{\ell, k_\ell}$ to its associated BS $\ell$.}
		\label{fig:system} \vspace{-0.05cm}
	\end{figure}
	{	As shown in Fig. \ref{fig:system}, we consider a cellular mmWave network consisting of $L$ cells. In each cell, a BS is deployed, and all the BSs coordinately serve the users in the network. Each BS transmits downlink (DL) ISAC signals to convey the data to users and receives the echoes to sense the states of users. In this work, the sensed information is utilized to assist the CSI acquisition to facilitate the coordination among the BSs. The details will be elaborated later.
		Denote the set of the BSs as $\mathcal{L}$. 
		Each BS is equipped with a uniform linear array (ULA) including $N_t$ transmit antennas and $N_r$ receive antennas, while each user is equipped with a single antenna. Denote the set of users associated with BS $\ell \in \Lc$  as $\Uc_\ell$, and $k_\ell$ as the $k$-th user. For coordinated transmission, the BSs are connected with each other via backhauls (e.g., IP-RAN \cite{xiaoming2020evolution}) of limited bandwidth. Consider a transmission duration including $N$ consecutive equal-length epochs and define $\Ec = \{1, \dots, N\}$. The length of each epoch is $\Delta T$. The transmission delay $T_d$ of the backhaul satisfies $T_d < \Delta T$.
		
		We consider block-fading propagation environments, and assume the DL channel $\hb_n^{\ell, k_\ell} \in \Cbb^{N_t}$ from BS $\ell\in \Lc$ to user $k_\ell$  keeps almost unchanged within the $n$-th epoch, while varying across different epochs. Since the BS is usually mounted at a high tower and its associated users are usually not far away from it, the LoS propagation path is highly achievable. Therefore, we assume that the LoS path exists in the intra-cell channels $\{\hb_n^{\ell, k_\ell}\}$ from each BS $\ell$ and its associated users $k_\ell \in \Uc_\ell$. In contrast, due to the long distance between BS $\ell$ to the users associated with the other BSs $m\neq\ell$, the LoS path is more likely to be blocked, thus we assume that the inter-cell channels $\{\hb_n^{m, k_\ell}\}_{m \neq \ell}$ does not have a LoS path for simplicity.  Denote $P_\ell > 0$ as the transmit power budget of BS $\ell \in \Lc$. The transmit signal for user $k_\ell$  in the $n$-th epoch is $s_n^{k_\ell}(t) \in \Cbb$, with $\vb_n^{k_\ell} \in \Cbb^{N_t}$ being its associated transmit BF. Besides, define the scheduling variable associated with user $k_\ell$ in the $n$-th epoch as 
		\vspace{-0.15cm}
		
		\begin{small} 
			\begin{equation}
				q_n^{k_\ell} \triangleq \left\{
				\begin{array}{l}
					1, ~\text{user $k_\ell$ is scheduled}, \\
					0,~ \text{otherwise}.
				\end{array}
				\right.
			\end{equation}
		\end{small}
		
		{\noindent}Based on these definitions, the DL transmit signal of BS $\ell\in \Lc$ in epoch-$n$ can be written as
		\vspace{-0.15cm}
		
		\begin{small}
			\begin{equation}
				\sbb_n^\ell(t) = \sum\nolimits_{k_\ell \in \Uc_\ell} q_n^{k_\ell} \vb_n^{k_\ell} s_n^{k_\ell}(t).
			\end{equation}
		\end{small}
		
		{\noindent}Accordingly, the received signal of user $k_\ell$ is
		\vspace{-0.15cm}
		
		\begin{small}
			\begin{equation}\label{eq:rx_sig}
				y_n^{k_\ell}(t) = \sum\nolimits_{m \in \Lc} (\hb_n^{m, k_\ell})^{\rm H} \sbb_n^\ell(t) + \omega_n^{k_\ell}(t),
			\end{equation} 
		\end{small} 
		
		{\noindent}where $\omega_n^{k_\ell}(t) \sim \mathcal{CN}(0, \sigma_c^2)$ is the additive white Gaussian noise (AWGN). Based on \eqref{eq:rx_sig}, the SINR of user $k_\ell$ in the $n$-th epoch $\gamma^{k_\ell}_{c,n }\left(\{q_n^{k_\ell}\},\{\vb_n^{k_\ell}\}; \{\hb^{\ell,k_\ell}_{n}\} \right)$ can be expressed as \eqref{eq:SINRc} at the top of next page, and the corresponding DL rate is
		\begin{small}
			\begin{figure*}[] 
				\centering
				\vspace{-0.0cm}
				\begin{equation} \label{eq:SINRc}
					\gamma_{c, n}^{k_\ell}\left(\{q_n^{k_\ell}\} ,\{\vb_n^{k_\ell}\}; \{\hb^{\ell,k_\ell}_{n}\} \right) = 
					\frac{q_{n}^{k_\ell} |(\hb^{\ell,k_\ell}_{n})^{\rm{H}} 
						\vb_{n}^{k_\ell} |^2}{ \underbrace{\sum\nolimits_{ i_\ell \neq k_\ell} q_{n}^{i_\ell} | (\hb^{\ell,k_\ell}_{n} )^{\rm{H}} \vb_{n}^{i_\ell} |^2}_{\text{intra-cell interference}} + \underbrace{\sum\nolimits_{m \neq \ell}\sum\nolimits_{j_m \in \Uc_m} q_{n}^{j_m} | (\hb^{m,k_\ell}_{n} )^{\rm{H}} \vb_{n}^{j_m} |^2}_{\text{inter-cell interference}} +  \sigma_{c}^2}
				\end{equation}
				\hrulefill \vspace{-0.4cm}
			\end{figure*}
		\end{small}
		
		\vspace{-0.3cm}
		
		\begin{small} 
			\begin{equation}\label{eq:rate}
				\begin{aligned} 
					R_{n}^{k_\ell} \big(\{q_n^{k_\ell}\} , &\{\vb_n^{k_\ell} \}; \{\hb^{\ell,k_\ell}_{n}\} \big)  \\
					& =\log_2 \left( 1 +\gamma_{c, n}^{k_\ell}\big(\{q_n^{k_\ell}\} ,\{\vb_n^{k_\ell}\}; \{\hb^{\ell,k_\ell}_{n}\} \big) \right).
				\end{aligned}
			\end{equation}	
		\end{small}
	}

	\vspace{-0.6cm}
	\subsection{Problem Description}
	\vspace{-0.1cm}
	{In this work, we are interested in the D-USCB design via the coordination among the BSs to improve the network throughput characterized by the PFR \cite{lee2013resource}, which is one of the most popular performance metrics for the transmission over consecutive epochs in practice. With \eqref{eq:rate}, the accumulative average rate of user $k_\ell$ is expressed as
		\vspace{-0.15cm}
		
		\begin{small}
			\begin{equation}
				\hat{R}_{N}^{k_\ell} \triangleq \frac{1}{N}\sum\nolimits_{n^\prime=1}^{N} R_{n^\prime}^{k_\ell} \big(\{q_{n^\prime}^{k_\ell}\} , \{\vb_{n^\prime}^{k_\ell} \}; \{\hb^{\ell,k_\ell}_{{n^\prime}}\} \big), 	
			\end{equation}
		\end{small}
		
		{\noindent}and the network PFR maximization problem can be formulated as 
		\vspace{-0.35cm}
		
		\begin{small}
			\begin{subequations} \label{p:PFR_SINR0}
				\begin{align}
					\!\!\!\!\!\!\max\limits_{\{q_n^{k_\ell}\} , \{\vb_n^{k_\ell} \}} &  \sum\nolimits_{\ell\in \mathcal{L}}  \sum\nolimits_{k_\ell \in 
						\mathcal{U}_\ell} \log\big( 	\hat{R}_{N}^{k_\ell} 	\big),  \label{eq:obj_PFR_SINR0}\\
					\text { s.t. } &q_{n}^{k_\ell} \in \{0,1\}, \forall k_\ell \in \mathcal{U}_\ell, \forall \ell \in \Lc, \forall n \in \Ec, \label{eq:cons_scheduling0} \\
					&\sum\nolimits_{k_\ell \in \mathcal{U}_\ell} q_{n}^{k_\ell} \|\vb_{n}^{k_\ell} \|^2 \leq P_\ell, \forall \ell \in \Lc, \forall n \in \Ec.  \label{eq:cons_pw0}			
				\end{align}       
			\end{subequations}
		\end{small}
		
		\vspace{-0.1cm}
		{\noindent}An common way to handle problem \eqref{p:PFR_SINR0} is solving the following problem at each epoch $n$ sequentially \cite{wu2021proportional}
		\vspace{-0.4cm}
		
		\begin{small} 
			\begin{subequations} \label{p:PFR_SINR}
				\begin{align}
					\! \! \! \max_{\{q_n^{k_\ell}\} , \{\vb_n^{k_\ell} \}} & \sum\limits_{\ell\in \mathcal{L}}  \sum\limits_{k_\ell \in 
						\mathcal{U}_\ell} {{R_{n}^{k_\ell}}\big(\{q_n^{k_\ell}\} , \{\vb_n^{k_\ell} \}; \{\hb^{\ell,k_\ell}_{n}\} \big)}/{{\hat{R}_{n-1}^{k_\ell}}},  \label{eq:obj_PFR_SINR}\\
					\text { s.t. } &q_{n}^{k_\ell} \in \{0,1\}, ~ \forall k_\ell \in \mathcal{U}_\ell, \forall \ell \in \Lc, \label{eq:cons_scheduling} \\
					&\sum\nolimits_{k_\ell \in \mathcal{U}_\ell} q_{n}^{k_\ell} \|\vb_{n}^{k_\ell} \|^2 \leq P_\ell,~ \forall \ell \in \Lc,  \label{eq:cons_pw}			
				\end{align}       
			\end{subequations} 
		\end{small}
		
		\vspace{-0.2cm}
		{\noindent}where ${\hat{R}_{n-1}^{k_\ell}}$ is the historical accumulative average rate and known in the $n$-th epoch.
		
		Problem \eqref{p:PFR_SINR} is difficult to solve, since it is non-convex and involves continuous variables and integer variables that are coupled in both the objective and constraints. More critically, the D-USCB optimization for \eqref{p:PFR_SINR} in practice faces great challenges, including
		\begin{enumerate}
			\item \textit{Inter-cell CSI acquisition}. Problem \eqref{p:PFR_SINR} relies on the availability of the intra-cell CSI $\{\hb_n^{\ell, k_\ell}\}$ and the (incoming) inter-cell CSI $\{\hb_n^{m, k_\ell}\}_{m \neq \ell}$. As discussed in Sec. \ref{sec:intro}, in the coordinated transmission mode of practical networks, BSs cannot obtain the CSI of non-served users but only the measured RSRP.

			\item \textit{High-dimension information exchange}. Even with available inter-cell CSI estimates $\{\hat\hb_{n}^{m, k_\ell}\}$, BSs $m\neq \ell$ need to exchange them to BS $\ell$ for the optimization in the latter. Exchanging the estimates of high-dimensional channels will lead to a large overhead in the backhaul.
		\end{enumerate}

		\vspace{-0.2cm}
		\section{Proposed SD-USCB Framework } \label{sec:proposed}	

        In this work, to overcome these challenges, we propose a novel SD-USCB framework by leveraging the ISAC transmission and CKM, as outlined in Fig. \ref{fig:D_USCB}. The proposed SD-USCB framework consists of the following key components: i) the combination of CKMs and proactive sensing for inter-cell CSI acquisition, ii) an SALINR-based approximate formulation to enable distributed optimization with reduced information exchange through backhauls, iii) the end-to-end procedure of the SD-USCB framework, including the ISAC signal transmission, the user scheduling,  the user location exchange, and the ISAC BF, and iv) the detailed algorithm design for the user scheduling and the ISAC BF. The details of our proposed designs are elaborated as follows.

		\vspace{-0.2cm}	
		\subsection{CSI acquisition by CKM and proactive sensing} \label{sec:CSI_acquisition} 
		
		To resolve the inter-cell CSI acquisition challenge, we propose integrating the CKM and the proactive sensing. 
		Specifically, each BS in the network is assumed to be equipped with a site-specific CKM, which can be constructed offline in advance through network measurement \cite{CKM_tut24_Zengyong, CKM_hyb_ZengyongTWC24}. The CKM establishes the one-to-one mappings from user locations in the respective coverage area of each BS to their CSI estimates. While CKMs with different types of CSI estimates have been developed, like the beam index map for hybrid BF \cite{CKM_hyb_ZengyongTWC24} and the clutter angle map for clutter suppression \cite{CKM_clutter_ICCC24}, they are for specific applications different from the D-USCB we consider here and thus can not be applied directly. In this work, the CKMs for the SD-USCB provide statistical CSI estimates $\{\hat\hb_n^{m, k_\ell}\}_{m\neq \ell}$ reflecting the large-scale propagation characteristics, which are extracted from the RSRP measurements by LSCM techniques \cite{zhang2023physics}. Details are illustrated in Sec. \ref{sec:CKM_cons}.

		With the constructed CKMs, the BS still needs to know the user locations to query their respective inter-cell CSI estimates. In the proposed SD-USCB, the locations are obtained by BSs via proactive sensing and exchanged to inform neighboring BSs. That is, each BS sends ISAC signals for simultaneous data communication and sensing. For the inter-cell channels $\{\hb_n^{m, k_\ell}\}_{m\neq \ell}$, the LoS path may not exist from the BS $m$ to the user $k_\ell$ associated with another BS $\ell$. Therefore, the location of user $k_\ell$ is sensed by its associated BS $\ell$ rather than BS $m$.  To attain $\{\hat\hb_n^{m, k_\ell}\}_{m\neq \ell}$, the location sensed by BS $\ell$ will be shared with BS $m$ via the backhaul linking them. Then, BS $m$ inputs this location into its CKM to obtain the CSI estimate.
		
		\vspace{-0.3cm}
		\subsection{SALINR to alleviate variable coupling and reduce the information exchange across BSs}  \label{sec:SLINR}
		
		\vspace{-0.1cm}
		While the inter-cell CSI estimates can be effectively acquired by constructing CKMs and proactive sensing, the D-USCB for \eqref{p:PFR_SINR} is still challenging due to the variable coupling across different BSs in the objective \eqref{eq:obj_PFR_SINR}.
		Specifically, the ICI term $\sum\nolimits_{m \neq \ell}\sum\nolimits_{j_m \in \Uc_m} q_{n}^{j_m} | (\hb^{m,k_\ell}_{n} )^{\rm{H}} \vb_{n}^{j_m} |^2 \triangleq \Ic_n^{k_\ell}$ in the SINR expression of ${R_{n}^{k_\ell}}(\cdot) $ is dependent on the scheduling variables $\{q_n^{j_m}\}$ and the BF variables $\{\vb_{n}^{j_m}\}$ of the other BSs $m \neq \ell$.   While conventional approaches \cite{tolli2009distributed,maros2017admm} based on primal/dual decomposition and ADMM enable distributed optimization, they require iterative updates and exchange of intermediate variables due to the coupling nature of problem \eqref{p:PFR_SINR}. Consequently, their overhead through the backhaul is substantial. To resolve it, we choose a different way by introducing a surrogate term, which can decompose the problem while avoiding iterative information exchange across BSs.

		As a surrogate, the introduced term should reflect the inter-cell interference level and not deviate too much from the true $\Ic_n^{k_\ell}$. In the existing literature \cite{sadek2007leakage,sadek2011leakage,tian2018new}, leakage-based metrics have been widely used to approximate inter-cell interference.  The latter reflects cross-cell interference from the user's perspective, while the former captures it from the BS's perspective. More importantly, the former has the merit of alleviating the variable coupling in the transmitter design. Specifically, conventional SLINR \cite{lopes2018leakage} approximates $\Ic_n^{k_\ell}$ by a leakage term defined as
			\begin{equation}
				\hat{L}_{n}^{k_\ell}=\sum\nolimits_{m\neq\ell}\sum\nolimits_{t_m\in\Uc_m} q_n^{t_m}\,
				\big|(\hb_n^{\ell,t_m}) \vb_n^{k_\ell}\big|^2,\label{leakage_beam}
			\end{equation}
			which can achieve a surrogate problem with fully decoupled BF variables across different BSs. 
            
        In this work, instead of applying SLINR, we introduce the averaged leakage from BS $\ell$ to the users in the other BSs to achieve a more accurate surrogate of $\Ic_n^{k_\ell}$, which is defined as the arithmetic mean  of the interference power generated by BS $\ell$ to the scheduled users in the other BSs \cite{cai2023approaching}, i.e.,
			\vspace{-0.3cm}
			
			\begin{small}\begin{equation}  \label{eq:leakage_avg}
					\begin{aligned}
						\tilde{L}_{n}^\ell & = 
						\sum\limits_{j_\ell \in \mathcal{U}_\ell}\sum\limits_{m \neq \ell}\sum\limits_{t_m \in \mathcal{U}_m}q_{n}^{j_\ell}{q_{n}^{t_m}}|
						{({\hb}^{\ell,t_m}_{n}\!)^{\rm{H}}} \vb_{n}^{j_\ell}|^2 /\sum\limits_{j_\ell \in \mathcal{U}_\ell}q_{n}^{j_\ell}, \\
						& = \sum\nolimits_{j_\ell\in\Uc_\ell}(q_n^{j_\ell} \hat{L}_{n}^{j_\ell})/{\sum\nolimits_{j_\ell\in\Uc_\ell} q_n^{j_\ell}}.
					\end{aligned}
				\end{equation} 
			\end{small}

			{\noindent}By replacing $\Ic_n^{k_\ell}$ in \eqref{eq:SINRc} with $\tilde{L}_{n}^\ell$, the SALINR, as an approximation of the original SINR, can be attained as
			
			\begin{small}\begin{equation} \label{eq:SALINR}
					\tilde{\gamma}_{n}^{k_\ell}
					=\frac{q_{n}^{k_\ell}|(\hb^{\ell,k_\ell}_{n} )^{\rm{H}} \vb_{n}^{k_\ell}|^2}{\sum_{i_\ell\neq k_\ell}q_{n}^{i_\ell} |(\hb^{\ell,k_\ell}_{n} )^{\rm{H}} \vb_{n}^{i_\ell} |^2+ \tilde{L}_{n}^\ell + \sigma_{c}^2}.
				\end{equation} 
			\end{small}
			
			\vspace{-0.2cm}
			{\noindent}The main difference of $\tilde{L}_{n}^\ell$ from the conventional leakage term  $\hat{L}_{n}^\ell$ \cite{li2022decentralized}  is the additional arithmetic mean operation over all the users $j_\ell \in \Uc_\ell$, thereby the optimization will consider the averaged leakage rather than the specific leakage. 
			
			To analyze the effectiveness of the SALINR-based approximation, denote the user rate as a function over the ICI as $R_n^{k_\ell}(\Ic_n^{k_\ell}) = \log\left(1 + {S_n^{k_\ell}}/
			({T_n^{k_\ell} + \Ic_n^{k_\ell} + \sigma_c^2})\right)$, where $S_n^{k_\ell}$ and $T_n^{k_\ell}$ correspond to the power of the received signal and the intra-cell interference, respectively. Accordingly, for the approximation errors between the true user rate $ R_n^{k_\ell}(\Ic_n^{k_\ell}) $ and the SLINR-based surrogate rate $R_n^{k_\ell} (\tilde{L}_{n}^{k_\ell})$,  as well as between $ R_n^{k_\ell}(\Ic_n^{k_\ell}) $ and the SALINR-based surrogate rate $  R_n^{k_\ell}(\hat{L}_{n}^{k_\ell}) $, the following theorem holds.
            \vspace{-0.4cm}
			\begin{Theorem} \label{Th:1}
				For any given set of beamformers $\{\vb_n^{k_\ell}\}_{k_\ell \in \Sc_\ell}$ with $\|\vb_n^{k_\ell}\| = P, \forall k_\ell \in \Sc_\ell, \ell \in \Lc$, under independent Rayleigh fading channels $\{\hb_n^{k_\ell}\}$, it holds that
				\vspace{-0.2cm}
				
				\begin{small}
					\begin{subequations} \notag
						\begin{align}
							\Ebb_{\{\hb_n^{k_\ell}\}} \left[\left| R_n^{k_\ell}(\tilde\Lc_n^{k_\ell}) - R_n^{k_\ell}(\Ic_n^{k_\ell})\right|\right] & \le  P \bar Z_n^{k_\ell}\sqrt{2 M_\ell}  ,\\
							\Ebb_{\{\hb_n^{k_\ell}\}} \left[\left| R_n^{k_\ell}(\hat\Lc_n^{k_\ell}) -R_n^{k_\ell}(\Ic_n^{k_\ell})\right|\right] & \le P \bar Z_n^{k_\ell}  \sqrt{M_\ell \big(1 + \frac{1}{|\Sc_n^\ell|} }\big),
						\end{align}
					\end{subequations}
				\end{small}
				
				\vspace{-0.4cm}
				{\noindent}where $\bar Z_n^{k_\ell}  >0$ is a parameter dependent on the Lipchitz constant of $R_n^{k_\ell}(\Ic_n^{k_\ell})$, $M_\ell=\sum_{m\neq\ell}\sum\nolimits_{t_m\in\Uc_m} q_n^{t_m}$, and  $\Sc_n^\ell$ is the set of scheduled users in cell $\ell$ with $|\Sc_n^\ell| = \sum\nolimits_{j_\ell\in\Uc_\ell}{q^{j_\ell}_n}$.
			\end{Theorem}
            \vspace{-0.5cm}
			\begin{Proof}
				The proof is detailed in Appendix \ref{App:proof}. 
			\end{Proof} 
			\vspace{-0.2cm}
			Theorem \ref{Th:1} indicates that SALINR-based rate approximation achieves a smaller error bound as long as multiple users are scheduled. Moreover, its bound decreases with an increased number of scheduled users. Therefore, SALINR can provide a more accurate approximation of the original SINR. It is indeed found that this SALINR-based optimization can attain a remarkable performance improvement compared with its SLINR-based counterpart, as shown in Sec. \ref{sec:simu_SALINR}.
			
			Based on SALINR, a surrogate rate function of $R_{n}^{k_\ell}(\cdot)$ can be obtained as $\tilde{R}_{n}^{k_\ell} \triangleq \log(1+\tilde{\gamma}_{n}^{k_\ell})$,  and problem \eqref{p:PFR_SINR} in each epoch $n$ is updated to 
		\vspace{-0.3cm}
		
		\begin{small}\begin{subequations} \label{p:PFR_SALINR}
				\begin{align}
					\max\nolimits_{\{\vb_{n}^{k_\ell},q_{n}^{k_\ell}\}} & \sum\nolimits_{\ell \in \mathcal{L}}\sum\nolimits_{k_\ell\in \mathcal{U}_{\ell}} {\tilde{R}_{n}^{k_\ell} }/{\hat{R}_{n-1}^{k_\ell}}, \label{eq:obj_PFR_SALINR}\\ 
					\text { s.t. } & \eqref{eq:cons_scheduling},\eqref{eq:cons_pw}.
				\end{align}  
			\end{subequations}
		\end{small}

		\vspace{-0.3cm}
		{\noindent}Compared with \eqref{p:PFR_SINR}, the BF variables $\{\vb_{n}^{k_\ell}\}$ from different BSs are successfully decoupled. While the scheduling variables are still coupled across BSs, they will be further handled by separating the user scheduling and BF optimizations into two stages \cite{li2015multicell}. 
		Additionally, as the inter-cell CSI required by each BS $\ell$ shifts from the incoming CSI in \eqref{eq:SINRc} to the outgoing CSI in \eqref{eq:leakage_avg}, cross-BS information exchange is simplified from “BS $\ell$ $\xrightarrow{\text{sensed locations of users $k_\ell$}}$ BS $m$ $\xrightarrow{\text{$\{\hat\hb_{n}^{m, k_\ell}\}$ queried from CKM}}$ BS $\ell$” to “BS $m$ $\xrightarrow{\text{sensed locations of users $t_m$}}$ BS $\ell$ (to query $\{\hat\hb_{n}^{\ell, t_m}\}$)”. The latter no longer needs to exchange high-dimensional inter-cell CSI estimates over the backhaul, significantly reducing the overhead.

		\vspace{-0cm}
		\subsection{Proposed SD-USCB procedure}
		\vspace{-0.1cm}
		After resolving the challenges of the inter-cell CSI acquisition and the information exchange, we further elaborate on the key steps of the SD-USCB framework in each epoch.
		
		Specifically, in every epoch, each BS $\ell$ sends beamformed ISAC signals to deliver information data to its scheduled users and sense their kinematic parameters, including the distance $d_n^{\ell, k_\ell}$, the angle $\theta_n^{\ell, k_\ell}$ and the velocity $v_n^{\ell, k_\ell}$. With these sensed parameters, the location of user $k_\ell$ can be estimated based on the geometric relation and further exchanged with neighboring BSs through the backhaul. The BSs then query their respective CKMs with the shared locations to obtain the (outgoing) inter-cell CSI, which is further used for the subsequent user scheduling and BF optimization.
		
		For ease of illustration, let us take the operations at BS $\ell$ in the $n$-th epoch within $[(n-1)\Delta T, n \Delta T]$ as an example. As shown in Fig. \ref{fig:framework}, the framework consists of four stages. 
		
		\begin{figure}[t]
			\centering
			{\resizebox{0.50\textwidth}{!}
				{\includegraphics{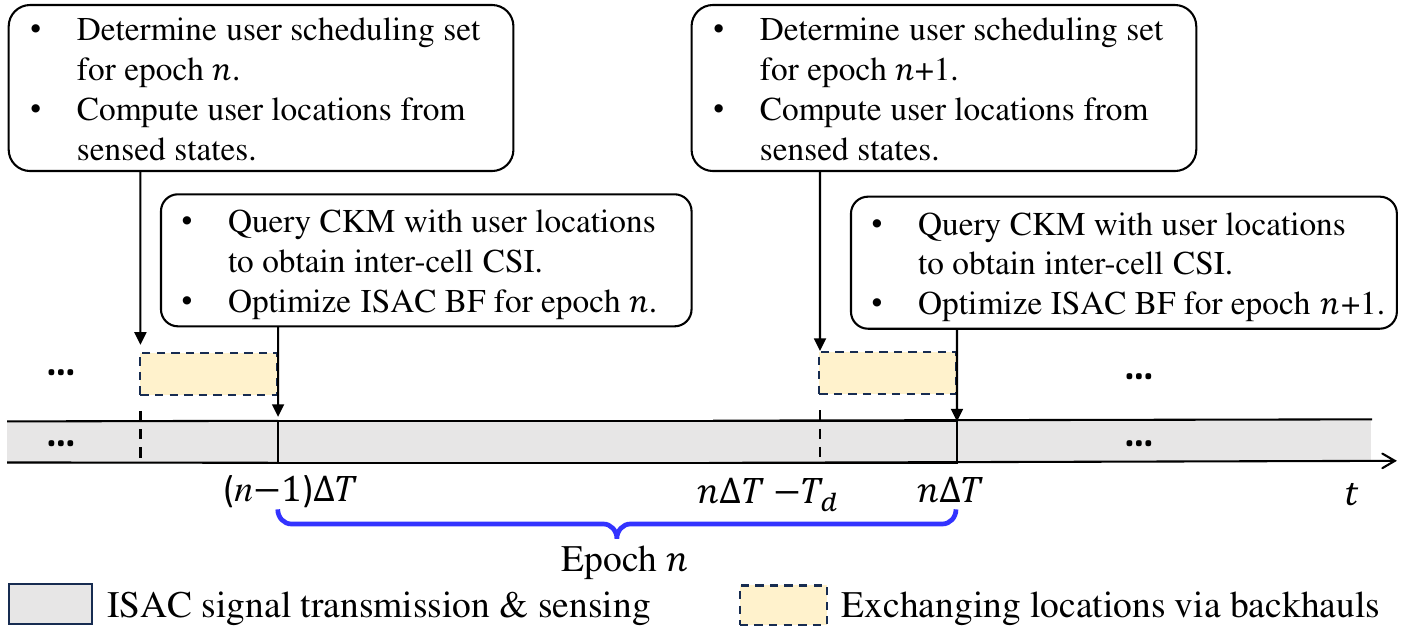}}}	\vspace{-0.5cm}
			\caption{ The operations of BS $\ell$ in the proposed SD-USCB framework.} 
			\label{fig:framework}\vspace{-0.1cm}
		\end{figure}
		
		\begin{itemize}
			\item {\bf Stage I}: \textit{ISAC Signal Transmission and Sensing}. At the beginning of the $n$-th epoch, $t = (n-1)\Delta T$, BS $\ell$ sends the beamformed signals $\sum_{k_\ell\in \mathcal{U}_\ell}\vb_n^{k_\ell} s_n^{k_\ell}(t)$. From the received echoes, BS $\ell$ acquires the estimates of the kinematic parameters $\{\hat{\theta}^{\ell,k_\ell}_{n},\hat{d}^{\ell,k_\ell}_{n},\hat{v}^{k_\ell}_{n}\}$ of served users.

			\item {\bf Stage II}: \textit{ Per-cell User Scheduling}. At $t = (n\Delta T- T_d)$, BS $\ell$ collects $\{\hat{R}_{n-1}^{k_\ell}\}_ {k_\ell \in \mathcal{U}_\ell}$ as the PFR weights of \eqref{eq:obj_PFR_SALINR}. With $\{\hat{R}_{n-1}^{k_\ell}\}_ {k_\ell \in \mathcal{U}_\ell}$ and the intra-cell CSI $\{\hat\hb_{n}^{\ell, k_\ell}\}_ {k_\ell \in \mathcal{U}_\ell}$, a user scheduling algorithm is conducted at BS $\ell$ to determine the scheduled user set $\Sc_{n}^\ell$ for epoch $n$.
			\item {\bf Stage III}: \textit{Information Exchange}. Based on $\Sc_{n}^\ell$, each BS $\ell$ computes the locations $\{(\hat{x}_{n}^{k_\ell}, \hat{y}_{n}^{k_\ell})\}_{k_\ell \in \Sc_n^\ell}$ from the sensed kinematic parameters and then exchanges them with other BSs $m \neq \ell$. Meanwhile, BS $\ell$ receives $\{(\hat{x}_{n}^{t_m}, \hat{y}_{n}^{t_m})\}_{t_m \in \Sc_n^m}$ and inquire its CKM to attain the inter-cell CSI $\{\hat{\hb}^{\ell,t_m}_{n}\}_{t_m \in \Sc_n^m}$.
			\item {\bf Stage IV}: \textit{Distributed ISAC BF}. At $t=n\Delta T$, with available intra-cell CSI $\{\hat\hb_{n}^{\ell, k_\ell}\}_ {k_\ell \in \Sc_n^\ell}$ and inter-cell CSI $\{\hat{\hb}^{\ell,t_m}_{n}\}_{t_m \in \Sc_n^m}$, each BS $\ell$ solves an SALINR-based PFR maximization problem to optimize the transmit ISAC BF $\{\vb_n^{k_\ell}\}$, which is applied to users in $\mathcal{S}_n^\ell$ for epoch $(n+1)$.
		\end{itemize}
		
		\vspace{-0.4cm}
		\subsection{Scheduling and BF Algorithm Design} \label{sec:alg_design}
		\vspace{-0.05cm}
		In this subsection, we illustrate the detailed designs for the four stages.
		
		{\bf ISAC Signal Transmission and Sensing}: Denote $\kappa = \sqrt{N_t N_r}$ as the gain of the antenna array and $\zb_{n-1}^\ell(t) \in \Cbb^{N_r}$ as the circularly symmetric complex Gaussian noise subject to $\zb_{n-1}^\ell(t)  \sim \mathcal{CN}(\mathbf{0}, \sigma^2 \Ib_{N_r})$. At an angle $\theta$ and with half-wavelength antenna spacing, the transmit steering vector and the receive steering vector are defined as 
		\vspace{-0.3cm}
		
		\begin{small}\begin{subequations} \label{eq:steer_trx}
				\begin{align}
					\ab(\theta) = \frac{1}{\sqrt{N_t}}[1, e^{j \pi \text{sin}(\theta)}, \!\dots, \! e^{j \pi (N_t-1)\text{sin}(\theta)}]^{\rm T} \in \Cbb^{N_t},\label{a_theta}\\
					\bb(\theta) = \frac{1}{\sqrt{N_r}}[1, e^{j \pi \text{sin}(\theta)}, \!\dots, \! e^{j \pi (N_r-1)\text{sin}(\theta)}]^{\rm T} \in \Cbb^{N_r},
				\end{align}
			\end{subequations}
		\end{small}
        
		\vspace{-0.2cm}
		{\noindent}respectively. Since the refraction power of the echo signals from other BSs is small, they are ignored for BS $\ell$.  Like \cite{LiuF_TWC20,Lei_JSAC25}, in the $(n-1)$-th epoch, the echoed ISAC signals received at BS $\ell$ can be expressed as
		\vspace{-0.4cm}
		
		\begin{small} \begin{equation} \label{eq:echo}
				\begin{aligned}
					\yb_{{\rm e}, n-1}^\ell(t)= & \sum\nolimits_{m \in\mathcal{L}} \sum\nolimits_{j_m\in \mathcal{U}_m}\!\kappa\beta^{\ell,j_m}_{n-1}	e^{j 2 \pi \mu^{\ell,j_m}_{n-1} t} \bb(\theta^{\ell,j_m}_{n-1}) \notag\\
					&\times \ab^{\rm{H}}(\theta^{\ell,j_m}_{n-1}){\sbb}^\ell_{n-1}(t-\tau^{\ell,j_m}_{n-1}) +\zb^{\ell}_{n-1}(t),
				\end{aligned}
			\end{equation}
		\end{small}
		
		{\noindent}where $\{\beta^{\ell,j_m}_{n-1}, \mu^{\ell,j_m}_{n-1},\tau^{\ell,j_m}_{n-1}\}$ are the reflection coefficient, the Doppler frequency shift and the round-trip delay of the sensing channel between BS $\ell$ and user $j_m$, respectively. Define the distance, angle, and velocity of user $k_\ell$ to BS $\ell$ in epoch $(n-1)$  as $d_{n-1}^{\ell, k_\ell}, \psi_{n-1}^{\ell, k_\ell}$ and $v_{n-1}^{\ell, k_\ell}$, respectively. According to the standard monostatic radar model~\cite{LiuF_TWC20}, the sensing parameters $\{\beta^{\ell,k_\ell}_{n-1}, \mu^{\ell,k_\ell}_{n-1}, \tau^{\ell,k_\ell}_{n-1}\}$ are deterministic functions of the kinematic parameters $\{d_{n-1}^{\ell,k_\ell}, \psi_{n-1}^{\ell,k_\ell}, v_{n-1}^{\ell,k_\ell}\}$.

		At BS $\ell$, the receive BF $\fb_{n-1}^{k_\ell} \triangleq \bb(\hat\theta^{\ell,k_\ell}_{n-2})$ is applied to enhance the echoes from user $k_\ell$, where $\hat\theta^{\ell,k_\ell}_{n-2}$ is the angle sensed in epoch $(n-2)$. Because the angular variation within a tiny interval across two consecutive epochs is small, with a large $N_r$ it holds that $|\bb^{\rm{H}}(\hat{\theta}_{n-2}^{\ell,k_\ell}) \bb({\theta}_{n-1}^{\ell,k_\ell})| \!\approx \! 1$, and $|\bb^{\rm{H}}(\hat{\theta}_{n-2}^{\ell,k_\ell}) \bb(\theta_{n-1}^{\ell,k^{\prime}_\ell})| \!=\! 0$ when $k^{\prime}_\ell \neq k_\ell$. Therefore, after receiving BF, the received echoed signal for user $k_\ell$ can be written as
		\vspace{-0.4cm}
		
		\begin{small}
			\begin{align}
				{r}_{{\rm e}, n-1}^{\ell,k_\ell}(t) & = (\fb_{n-1}^{k_\ell})^{\rm H}\yb_{{\rm e}, n-1}^\ell(t),\label{eq:echo_after_BF}\\
				&~\begin{aligned}
					\approx &~\kappa\beta^{\ell,k_\ell}_{n-1}	e^{j 2 \pi\mu^{\ell,k_\ell}_{n-1} t} \ab^{\rm{H}} (\theta_{n-1}^{\ell,k_\ell} )\sbb_{n-1}^\ell (t - \tau_{n-1}^{\ell,k_\ell} ) + {z}_{ n-1}^{\ell,k_\ell} (t)\notag,
				\end{aligned} 
			\end{align}
		\end{small}
		
		\vspace{-0.15cm}
		{\noindent}where ${z}_{n-1}^{\ell,k_\ell}(t)\!\sim\! \mathcal{CN}(0,\sigma_z^2)$. Based on ${r}_{{\rm e}, n-1}^{\ell,k_\ell}(t)$, BS $\ell$ can adopt the matched filtering to estimate the delay $\tau_{n-1}^{\ell,k_\ell}$ and the Doppler frequency shift $\mu^{\ell,k_\ell}_{n-1}$, then the angle $\theta_{n-1}^{\ell,k_\ell}$ can be attained by maximum-likelihood estimation. The resulting measurements are modeled by additive errors $z_{ n-1}^{\ell,k_\ell,i},\, i \in \{\tau, \mu, \theta\}$, are assumed to be zero-mean complex Gaussian random variables, i.e., $z_{ n-1}^{\ell,k_\ell,i} \sim \mathcal{CN}(0,\sigma_{k_\ell,i}^2)$, where the variance $\sigma_{k_\ell,i}^2$ is inversely proportional to the sensing SINR of user~$k_\ell$ \cite{LiuF_TWC20, Lei_JSAC25}, and can be explicitly expressed as
		\vspace{-0.4cm}
		
		\begin{small}\begin{equation} \label{eq:SINRs}
				\!\!\!\sigma_{k_\ell,i}^2\!=\!\frac{a_i^2 (\sum_{j_\ell\neq k_\ell} 	\!	G\kappa^2 	q_{n-1}^{j_\ell} |\beta_{n-1}^{\ell,k_\ell} |^2 |\ab^{ {\rm{H}}}( \theta_{n-1}^{\ell,k_\ell}) \vb^{j_\ell}_{n-1}|^2+\sigma_z^2)}{G \kappa^2 |\beta_{n-1}^{\ell,k_\ell} |^2 |\ab^{\rm{H}}(\theta_{n-1}^{\ell,k_\ell}) \vb^{k_\ell}_{n-1} |^2}, 
			\end{equation}
		\end{small}
		
		{\noindent}where $G$ is the matched-filtering gain, and the parameters $a_i,  i \in  \{\tau, \mu, \theta\}$ are determined according to the system configuration and specific parameter estimation algorithms.

		{\bf Per-cell Scheduling}: Notice that the scheduling variables $\{q_{n}^{k_\ell}\}$ and the BF variables $\{\vb_{n}^{k_\ell}\}$ are still coupled in problem \eqref{p:PFR_SALINR}. {Moreover, only the intra-cell CSI $\{\hb_n^{\ell, k_\ell}\}$ is available at each BS $\ell$, while the scheduling of other cells  $\{q_n^{j_m}|q_n^{j_m}=1\}_{ m \in \Lc, m\neq\ell}$ and the inter-cell CSI $\{\hat \hb_n^{\ell, j_m}\}_{m \neq \ell}$ remain unknown,} making the joint scheduling among the BSs difficult.{
		Therefore, we first address the intra-cell interference by per-cell user scheduling, leaving the ICI alleviation to the subsequent ISAC BF stage.   Each BS performs scheduling greedily using only $\{\hat{\hb}_n^{\ell, k_\ell}\}$ to maximize the PFR.
            
        The scheduling design prioritizes two criteria: i) users not scheduled for a long time (e.g., in the previous $T_s$ epochs) should be immediately scheduled to enhance the PFR; ii) spatially separated users with weak channel correlation are favored.  This mitigates the intra-cell interference in \eqref{eq:SINRc} in the DL transmission while alleviating the beam overlap, thereby enhancing the sensing in \eqref{eq:SINRs}. 
         
		{Following these ideas, we initialize the scheduling set $\Sc_{n}^\ell$ with users $\mathcal{K}_{s} \!\! =\!\! \{\pi_1, \ldots,\pi_{j-1}\}$ unscheduled for the past $T_s -1$  epochs.  Let $\bar\Sc_n^\ell = \Uc_\ell \backslash \Sc_n^\ell$ contain the remaining unscheduled users. We then iteratively expand $\Sc_{n}^\ell$: each user in $\bar\Sc_n^\ell$ is tentatively added, and ZF BF with equal power allocation is applied to the expanded set to compute the PFR.
        The user whose addition yields the highest PFR is permanently added to $\Sc_n^\ell$. 
        This process is repeated until the PFR can no longer be improved by adding any unscheduled user. Obviously, the users are selected in a greedy manner according to the PFR metric. The full procedure of the proposed PFZFG scheduling at BS $\ell$ in the $n$-th epoch is outlined in Algorithm \ref{alg:PFZFG}.
		}
		
		\begin{figure}
			\begin{algorithm}[H]
				\begin{small}
					\caption{PFZFG User Scheduling Algorithm.}
					\begin{algorithmic}[1] \label{alg:PFZFG}
						\STATE {\bf Given} $\Kc_{s}, \Sc_n^\ell = \emptyset$;
						\STATE Initialize $\Sc_n^\ell \leftarrow \Kc_{s}$, and compute the set of the remaining users set  by $\bar\Sc_n^\ell = \Uc_\ell \backslash \Sc_n^\ell$;
						\STATE Compute $R' = \!\sum_{i \in \Kc_s} \! R^{i}_{n}$ based on ZF-BF with equal power.
						\WHILE {True}
						\STATE Compute 		
						$\pi_j = \arg \max\limits_{ k \in \bar\Sc_n^\ell} \sum\limits_{i\in \Sc_n^{\ell,k}} {R^{i}_{n} }/{\hat{R}^i_{n-1}}$, where $\Sc_n^{\ell,k} \leftarrow \Sc_n^\ell \cup k, \forall k \in \bar\Sc_n^\ell$;
						\STATE Compute $R = \sum\nolimits_{i\in \Sc_n^{\ell,\pi_j}} {R^i_{n} }/{\hat{R}^i_{n-1}}$;
						\IF {$R\geq R^{\prime}$}
						\STATE Update  $\Sc^\ell_n  \leftarrow \Sc^\ell_n \cup \pi_j, $ and $\bar\Sc_n^\ell \leftarrow \bar\Sc_n^\ell \backslash \pi_j$;
						\STATE Update $ R' \leftarrow R$, and $j \leftarrow j+1$;  
						\ELSE
						\STATE Break;
						\ENDIF
						\ENDWHILE
						\STATE {\bf Output} the set of the scheduled users $\Sc_n^\ell$.
						
					\end{algorithmic} 
				\end{small}
			\end{algorithm}
		\end{figure}
		
		{\bf Information Exchange}: After determining  $\Sc_n^\ell$, each BS $\ell \in \Lc$ will compute the locations $\{(\hat x_n^{k_\ell}, \hat y_n^{k_\ell})\}_{k_\ell \in \Sc_{n}^\ell}$ from the sensed kinematic parameters based on the geometric relation \cite{LiuF_TWC20}. BS $\ell$ then forwards the locations $\{(\hat x_n^{k_\ell}, \hat y_n^{k_\ell})\}_{k_\ell \in \Sc_{n}^\ell}$ to the other BSs $m \neq \ell$ and receives their  scheduled users' locations.  With the received locations, BS $\ell$ queries its pre-constructed CKM to attain the inter-cell CSI estimates $\{\hat\hb_{n}^{\ell, j_m}\}_{j_m \in \Sc_{n}^m}$.
		
		{\bf Distributed ISAC BF}: Given the scheduling set $\Sc_{n}^\ell$ and the inter-cell CSI $\{\hat\hb_{n}^{\ell, j_m}\}_{j_m \in \Sc_{n}^m}$, variable coupling across BSs in \eqref{p:PFR_SALINR} is eliminated, allowing each BS to optimize its beamformers independently. The beamformer must serve both communication and proactive sensing, where sensing performance affects future CSI acquisition and BF design. Therefore, \eqref{p:PFR_SALINR} should account for sensing quality. To this end, we constrain the variance of the proactive sensing errors to stay below a threshold $\bar c$. At each BS $\ell$, the beamformers are then optimized by solving the following virtual PFR maximization problem, subject to the transmit power constraint and sensing error constraints	
		\vspace{-0.3cm}
		
		\begin{small} \begin{subequations}\label{p:BF_ISAC_opt}
				\vspace{-0.1cm}
				\begin{align}
					{\max\nolimits_{\{\vb_{n}^{k_\ell}\}}} 
					~&\sum\nolimits_{k_\ell \in\mathcal{S}_n^\ell} {\tilde{R}^{k_\ell	}_{n}}/{\hat{R}_{n-1}^{k_\ell}}, \label{eq:obj_ISACBF}\\
					\text {s.t.} 	
					& \sum \nolimits_{k_\ell \in\mathcal{S}_n^\ell} \|\vb^{k_\ell}_{n}\|^2 \leq P_\ell, \label{eq:cons_pw_ISACBF}\\
					&\sigma_{k_\ell,i}^2	( \{\vb_{n}^{k_\ell}\} )\leq \bar c, i \in  \{\tau, \mu, \theta\}, k_\ell \in \mathcal{S}_n^\ell. \label{eq:cons_crb}
					\vspace{-1.5cm}
				\end{align}
			\end{subequations}
		\end{small}
		
		\vspace{-0.2cm}
		{\noindent}Because of the objective function \eqref{eq:obj_ISACBF} and the sensing constraints \eqref{eq:cons_crb}, problem \eqref{p:BF_ISAC_opt} is non-convex. To solve it, we first cope with the non-convexity of \eqref{eq:obj_ISACBF} by FP techniques \cite{shen2019optimization}. {Introducing auxiliary variables $\{\xi_n^{k_\ell}\}$ and leveraging the Lagrangian dual transform \cite{rostami2017admm}, \eqref{eq:obj_ISACBF} is equivalent to }
		\vspace{-0.4cm}
		
		\begin{small}\begin{equation} \label{p:BF_ISAC_opt_kesi}
				\begin{aligned}
					\!\!\max_{\{\vb_{n}^{k_\ell}, \}, \{\xi_n^{k_\ell}\}} \sum\nolimits_{k_\ell \in\Sc_n^\ell}  &  \tilde{R}_{n}^{k_\ell}/\hat{R}_{n-1}^{k_\ell}-\xi_n^{k_\ell}/\hat{R}_{n-1}^{k_\ell} \\ 
					+ \!\sum\nolimits_{k_\ell \in\Sc_n^\ell}  & \frac{{ (1+\xi_n^{k_\ell})/\hat{R}_{n-1}^{k_\ell}} \Re((\hat{\hb}^{\ell,k_\ell}_{n})^{\rm{H}} \vb^{k_\ell}_{n})}{\sum\nolimits_{s_\ell \in\Sc_n^{\ell}}  | 	(\hat{\hb}^{\ell,k_\ell}_{n})^{\rm{H}}  \vb_{n}^{s_\ell} |^2 \! + \!{ L_{n}^\ell +\sigma_c^2}}	.
				\end{aligned}
			\end{equation}	
		\end{small}
		
		\vspace{-0.2cm}
		{\noindent}With fixed $\{\vb_{n}^{ k_\ell}\}$, the optimal values of $\{\xi_n^{k_\ell}\}$ can be attained based on the first-optimal condition in a closed-form as
		\vspace{-0.2cm}
		
		\begin{small}
			\begin{equation}\label{xi}
				(\xi_n^{k_\ell})^* = \tilde\gamma_{n}^{k_\ell}, \forall k_\ell \in \Sc^\ell_n.
			\end{equation} 
		\end{small}
		
		\vspace{-0.1cm}
		{\noindent}Given $\{\xi_n^{k_\ell}\}$, the objective function in \eqref{p:BF_ISAC_opt_kesi} is still a sum-of-ratio form. To handle it, by further introducing auxiliary variables $\{\zeta_n^{k_\ell}\}$, \eqref{p:BF_ISAC_opt_kesi} is equivalent to 
		\vspace{-0.35cm}
		
		\begin{small} \begin{equation}\label{eq:obj_FP}
				\max_{\{\vb_{n}^{k_\ell} \}, \{\xi_n^{k_\ell}\}, \{\zeta_n^{k_\ell}\}}~\sum\nolimits_{k_\ell \in\Sc_n^\ell} f_{k_\ell}(\{\vb_{n}^{k_\ell}\}, \{\xi_n^{k_\ell}\}, \{\zeta_n^{k_\ell}\}),  
			\end{equation}
		\end{small}
		
		\vspace{-0.15cm}
		{\noindent}where 
		\vspace{-0.4cm}
		
		\begin{small}
			\begin{equation}\label{kesi}
				\begin{aligned}
					f_{k_\ell}&(\{\vb_{n}^{k_\ell} \}, \{\xi_n^{k_\ell}\}, \{\zeta_n^{k_\ell}\}) = \\
					&\sum\nolimits_{k_\ell \in\Sc_n^\ell}  \bigg\{ 2 \zeta_n^{k_\ell} \!	\sqrt{ (1+\xi_n^{k_\ell})/\hat{R}_{n-1}^{k_\ell}} \Re(({\hb}^{\ell,k_\ell}_{n})^{\rm{H}} \vb^{k_\ell}_{n}) \\
					& -
					(\zeta_n^{k_\ell})^2\bigg[
					\sum\nolimits_{s_\ell \in\Sc_n^{\ell}}  | 	(\hat{\hb}^{\ell,k_\ell}_{n})^{\rm{H}}  \vb_{n}^{s_\ell} |^2 \! + \!{ L_{n}^\ell +\sigma_c^2}
					\bigg]\! \bigg\}.
				\end{aligned}
			\end{equation}
		\end{small}
		
		\vspace{-0.2cm}
		{\noindent}With fixed $\{\vb_{n}^{k_\ell}, \xi_n^{k_\ell}\}$, $\{\zeta_n^{k_\ell}\}$ admit closed-form solutions 
		\vspace{-0.1cm}
		
		\begin{small}
			\begin{equation}
				(\zeta_n^{k_\ell})^*= \frac{\sqrt{{ (1+\tilde{\gamma}^{k_\ell}_n )}|(\hat{\hb}^{\ell,k_\ell}_{n})^{\rm{H}} \vb_{n}^{k_\ell} |^2 /{\hat{R}_{n-1}^{k_\ell}}}}{\sum_{j_\ell\in \Sc_n^\ell}  |(\hat{\hb}^{\ell,k_\ell}_{n})^{\rm{H}} \vb_{n}^{j_\ell} |^2 + L_{n}^\ell + \sigma_c^2}.	
			\end{equation}
		\end{small}
		
		\vspace{-0.1cm}
		{\noindent}In addition, denote $\bar{i} = \argmax_i a_i$, the sensing constraints \eqref{eq:cons_crb} are equivalent to
		\vspace{-0.3cm}
		
		\begin{small}\begin{equation}\begin{aligned}
					\label{eq:cons_crb1}
					\phi_{k_\ell}(\{\vb_n^{j_\ell}\}) \leq \psi_{k_\ell}^{\bar{i}}(\vb_n^{k_\ell}),  \forall k_\ell, j_\ell
				\end{aligned}
			\end{equation}
		\end{small}
        
        \vspace{-0.1cm}
		{\noindent}where $
                \phi_{k_\ell}(\{\vb_n^{j_\ell}\})
                \triangleq
                \sum_{j_\ell \in \Sc_n^\ell, j_\ell \neq k_\ell}
                G \kappa^2 |\hat{\beta}_{n}^{\ell,k_\ell}|^2 
                | \ab^{\rm H}(\hat{\theta}_{n}^{\ell,k_\ell}) \vb_n^{j_\ell}|^2$ $+ \sigma_z^2,
                \psi_{k_\ell}^{\bar{i}}(\vb_n^{k_\ell})
                \triangleq
                {\bar{c} G \kappa^2 |\hat{\beta}_{n}^{\ell,k_\ell}|^2}/{a_i^2}
                | \ab^{\rm H}(\hat{\theta}_{n}^{\ell,k_\ell}) \vb_n^{k_\ell}|^2$,
        and
		 $\hat{\beta}_{n}^{\ell,k_\ell} = \eta^{k_\ell}/(\hat{\tau}_{n}^{\ell,k_\ell}c)$.

        Based on \eqref{eq:obj_FP} and \eqref{eq:cons_crb1}, problem \eqref{p:BF_ISAC_opt} is transformed into 
		\vspace{-0.3cm}
		
		\begin{small}\begin{subequations} \label{p:BF_ISAC_opt1}
				\begin{align}
					\max_{\{\vb_{n}^{k_\ell}\}, \{\xi_n^{k_\ell}\}, \{\zeta_n^{k_\ell}\}}&~\sum\nolimits_{k_\ell \in\Sc_n^\ell} f_{k_\ell}(\{\vb_{n}^{k_\ell} \}, \{\xi_n^{k_\ell}\}, \{\zeta_n^{k_\ell}\}), \\
					\text {s.t.} &	~\eqref{eq:cons_pw_ISACBF},\eqref{eq:cons_crb1},
				\end{align}
			\end{subequations}
		\end{small}
		
		\vspace{-0.2cm}
		{\noindent}which can be solved by updating the three types of variables sequentially in a block coordinate descent (BCD) manner.

        However, when fixing $\{\xi_n^{k_\ell}\}$ and $\{\zeta_n^{k_\ell}\}$, the non-convex sensing constraints~\eqref{eq:cons_crb1} render the subproblem w.r.t.\ $\{\vb_{n}^{k_\ell}\}$ difficult to solve. A possible approach~\cite{SD_USBF_icassp24} is to apply SDR, which converts the subproblem into a convex semidefinite program (SDP), but at the cost of high-dimensional matrix variables and increased complexity. To improve efficiency, we instead solve~\eqref{p:BF_ISAC_opt1} directly over the BF vector variables.

        First, the sensing constraints are convexified by SCA. Given a fixed beamformer $\vb_{n}^{k_\ell,(r)}$ with $r$ being the iteration index, a lower bound of $\psi_{k_\ell}^{\bar{i}}(\vb_n^{k_\ell})$ in \eqref{eq:cons_crb1} can be obtained via the first-order Taylor expansion as

        \vspace{-0.4cm}
        \begin{small}
            \begin{align}\label{eq:linear_bound}
                & \tilde{\psi}_{k_\ell}^{\bar{i}}(\vb_n^{k_\ell}|\vb_{n}^{k_\ell,(r)})
                \notag \\ 
                &=  2 \Re \{
                (\vb_{n}^{k_\ell,(r)})^{\rm H} \Cb_n^{k_\ell} \vb_n^{k_\ell}
                \}
                \!-\!
                (\vb_{n}^{k_\ell,(r)})^{\rm H} \Cb_n^{k_\ell} \vb_{n}^{k_\ell,(r)},
            \end{align}
        \end{small}

        \vspace{-0.2cm}
        \noindent where {\small $\Cb_n^{k_\ell}
        =
        \bar{c} G \kappa^2 |\hat{\beta}_{n}^{\ell,k_\ell}|^2
        \ab(\hat{\theta}_{n}^{\ell,k_\ell})
        \ab^{\rm H}(\hat{\theta}_{n}^{\ell,k_\ell})/{a_i^2}.$}
        By replacing {\small$\psi_{k_\ell}^{\bar{i}}(\vb_n^{k_\ell})$} with {\small$\tilde{\psi}^{\bar{i}}_{k_\ell}(\vb_n^{k_\ell}|\vb_{n}^{k_\ell,(r)})$},  the subproblem of \eqref{p:BF_ISAC_opt1} w.r.t. {\small$\{\vb_{n}^{k_\ell}\}$} is approximated as the following convex problem

        \vspace{-0.4cm}
        \begin{small}
            \begin{subequations}        \label{p:BF_ISAC_opt2}
                \begin{align}
                    \max\nolimits_{\{\vb_{n}^{k_\ell}\}}
                    &~
                    \sum\nolimits_{k_\ell \in\Sc_n^\ell}
                    f_{k_\ell}(\{\vb_{n}^{k_\ell}\}; \{\xi_n^{k_\ell}\}, \{\zeta_n^{k_\ell}\}),
                    \\
                    \text{s.t.}\quad
                    &
                    \sum\nolimits_{k_\ell \in\Sc_n^\ell} \|\vb^{k_\ell}_{n}\|^2 \le P_\ell,
                    \label{eq:cons_power_app}
                    \\
                    &
                    \phi_{k_\ell}(\{\vb_n^{j_\ell}\})
                    \le
                    \tilde{\psi}_{k_\ell}^{\bar{i}}(\vb_n^{k_\ell}|\vb_{n}^{k_\ell,(r)}),
                    \quad \forall k_\ell \in \Sc_n^\ell.
                    \label{eq:cons_sensing_app}
                \end{align}
            \end{subequations}
        \end{small}
        \vspace{-0.4cm}
        
        \noindent While \eqref{p:BF_ISAC_opt2} can be solved by off-the-shelf solvers like CVX, a computation-efficient algorithm is more appealing for practical deployment. Therefore, we further develop a DualOpt-based first-order algorithm.

        Specifically, introducing dual variables $\lambda \ge 0$ for the power constraint \eqref{eq:cons_power_app} and $\{\mu_{k_\ell} \ge 0\}$ for the sensing constraints \eqref{eq:cons_sensing_app}, the Lagrangian of \eqref{p:BF_ISAC_opt2} is given by

        \vspace{-0.2cm}
        \begin{small}
            \begin{align}\label{eq:lagrangian}
                &\mathcal{L}(\{\vb_{n}^{k_\ell}\}, \lambda, \{\mu_{k_\ell}\})
                =
                -
                \lambda
                \sum\nolimits_{k_\ell \in \Sc_n^\ell}
                \|\vb_n^{k_\ell}\|^2
                +
                \lambda P_\ell \notag\\
                  & + \sum\nolimits_{k_\ell \in \Sc_n^\ell}
                \Bigl(
                2 \Re \{ (\zb^{k_\ell}_n)^{\rm H} \vb_n^{k_\ell} \}
                -
                (\vb_n^{k_\ell})^{\rm H} \Ab^{k_\ell}_n \vb_n^{k_\ell}
                \Bigr),
            \end{align}
        \end{small}
        \vspace{-0.2cm}
        
        \noindent where $\zb^{k_\ell}_n
        \triangleq
        \zeta_n^{k_\ell}
        \sqrt{{1+\xi_n^{k_\ell}}/{\hat{R}_{n-1}^{k_\ell}}}
        \hat{\hb}_n^{\ell,k_\ell}
        +
        \mu_{k_\ell} \Cb_n^{k_\ell} \vb_n^{k_\ell,(r)},$ $\Ab_n^{k_\ell}
        \triangleq
        \sum_{j_\ell \in \Sc_n^\ell}
        (\zeta_n^{j_\ell})^2
        \hat{\hb}_n^{\ell,j_\ell}
        (\hat{\hb}_n^{\ell,j_\ell})^{\rm H}  +
        ( \sum_{j_\ell \in \Sc_n^\ell} (\zeta_n^{j_\ell})^2 
        )\Db^\ell_n
        +
        \mu_{k_\ell} \Cb_n^{k_\ell},$
        and $\Db_n^\ell $ denotes the leakage covariance matrix
        
        \vspace{-0.2cm}
        \begin{small}
            \begin{equation}
                \Db^\ell_n
                = \sum\nolimits_{m \neq \ell}\sum\nolimits_{t_m \in \mathcal{U}_m}
                 {q_{n}^{t_m} \hat{\hb}^{\ell,t_m}_{n}(\hat{\hb}^{\ell,t_m}_{n})^{\rm{H}}}
                \big/{\sum\nolimits_{j_\ell \in \mathcal{U}_\ell}q_{n}^{j_\ell}}.
            \end{equation}
        \end{small}
        
        \vspace{-0.2cm}
        \noindent By setting $\nabla_{\vb_n^{k_\ell}} \mathcal{L}(\cdot) = \mathbf{0}$, the optimal beamformer admits the following semi-closed-form expression

        \vspace{-0.2cm}
        \begin{small}
            \begin{equation}\label{upadte-v}
                \vb_n^{k_\ell,(r,s+1)}
                =
                (
                \Ab^{k_\ell}_n
                +
                \lambda^{(r,s)} \Ib
                )^{-1}
                \zb^{k_\ell}_n,
                \quad \forall k_\ell \in \Sc_n^\ell.
        \end{equation}
        \end{small}
        
        \vspace{-0.2cm}
        \noindent Substituting \eqref{upadte-v} into \eqref{eq:lagrangian}, the dual problem of \eqref{p:BF_ISAC_opt2} is 

        \vspace{-0.3cm}
        \begin{small}
             \begin{subequations} \label{p:dual}
        	\begin{align}
        		\min\nolimits_{ \lambda, \{\mu_{k_\ell}\}} & \ g(\lambda, \{\mu_{k_\ell}\})  \\
        		\text{s.t.} & \ \lambda \ge 0, \mu_{k_\ell} \ge 0, \forall k_\ell \in \Sc_n^\ell
        	\end{align}
        \end{subequations}
        \end{small}
        
        \vspace{-0.2cm}
        \noindent with $ g(\lambda,\{\mu_{k_\ell}\})
        =
        \sum_{k_\ell \in \Sc_n^\ell}
        (\zb_n^{k_\ell})^{\rm H}
        \bigl(
        \Ab_n^{k_\ell}
        +
        \lambda \Ib
        \bigr)^{-1}
        \zb_n^{k_\ell}
        +
        \lambda P_\ell $. 
        
        Now \eqref{p:dual} is a convex problem with simple linear inequalities, so it can be efficiently solved by projected gradient descent (PGD) method. Specifically, in each dual iteration $(s+1)$, the dual variables can be updated as 

        \vspace{-0.2cm}
        \begin{small}
            \begin{subequations} \label{eq:dual_opt}
        	   \begin{align}
                    &\!\!\lambda^{(r,s+1)}
                     =
                    \Bigl[
                    \lambda^{(r,s)}
                    \!-\!
                    \alpha_\lambda
                    \Bigl( P_\ell \!-\!
                    \sum\nolimits_{k_\ell \in \mathcal{S}_n^\ell}
                    \|\mathbf{v}_n^{k_\ell}\|^2
                    \Bigr)
                    \Bigl]^+, \\
                    \!\!&\!\!\!\mu_{k_\ell}^{(r,s+1)}
                    \!=\!
                    \Bigl[
                    \mu_{k_\ell}^{(r,s)}
                    \!-\!
                    \alpha_\mu
                    \Bigl(
                    \tilde{\psi}_{k_\ell}^{\bar{i}}
                    (\mathbf{v}_n^{k_\ell}\!\mid\!\mathbf{v}_n^{k_\ell,(r,s)})\!-\!\phi_{k_\ell}(\{\mathbf{v}_n^{j_\ell}\})
                    \Bigr)
                    \Bigr]^+\!\!,
                \end{align}
            \end{subequations}
        \end{small}
    
        \noindent where $\alpha_{\lambda }$ and $\alpha_{\mu}$ are non-negative step sizes, and $[\cdot]^+$ denotes the projection onto the nonnegative orthant. After solving problem \eqref{p:dual} by the dual updates in \eqref{eq:dual_opt}, the reference points of SCA and FP intermediate variables can be updated in an outer loop. Accordingly, the proposed DualOpt-Based ISAC BF for \eqref{p:BF_ISAC_opt} is summarized in Algorithm \ref{alg:BF_FPSCALD}. 
        
        \begin{figure}[t]
        \vspace{-0.25cm}
        \begin{algorithm}[H]
        \caption{Proposed DualOpt-Based ISAC BF Algorithm for Problem \eqref{p:BF_ISAC_opt}}
        \label{alg:BF_FPSCALD}
        \begin{algorithmic}[1]
        
        \STATE \textbf{Initialization:} Initialize beamformers $\{\vb_n^{k_\ell,(0)}\}$, 
        dual variables $\lambda^{(0)}$ and $\{\mu_{k_\ell}^{(0)}\}$, 
        step sizes $\{\alpha_{\lambda}, \alpha_{\mu_{k_\ell}}\}$, 
        and set outer iteration index $r=0$.
        
        \REPEAT
            \STATE \textbf{(FP update)} Update auxiliary variables 
            $\{\xi_n^{k_\ell,(r)}\}$ via \eqref{xi} and 
            $\{\zeta_n^{k_\ell,(r)}\}$ via \eqref{kesi}.
            
            \STATE Initialize inner iteration index $s=0$ and 
            $\vb_n^{k_\ell,(r,0)} = \vb_n^{k_\ell,(r)}$.
            
            \REPEAT 
                \STATE Compute $\{\vb_n^{k_\ell,(r,s)}\}$ by \eqref{upadte-v}.
                
                \STATE Update $\{\mu_{k_\ell}^{(r,s)}\}$ and $\lambda^{(r,s)}$ by \eqref{eq:dual_opt}.
                
                \STATE $s \leftarrow s+1$.
            \UNTIL{dual updates converge.}
            
            \STATE Set $\vb_n^{k_\ell,(r+1)} = \vb_n^{k_\ell,(r,s)}$, 
            $\lambda^{(r+1)} = \lambda^{(r,s)}$, 
            $\{\mu_{k_\ell}^{(r+1)}\} = \{\mu_{k_\ell}^{(r,s)}\}$.
            
            \STATE $r \leftarrow r+1$.
            
        \UNTIL{outer FP iteration converges.}
        
        \STATE \textbf{Output:} Optimized beamformers $\{\vb_n^{k_\ell, (r)}\}$.
        
        \end{algorithmic}
        \vspace{-0.15cm}
        \end{algorithm}
        \end{figure}

        It is worth noting that the main computations of Algorithm \ref{alg:BF_FPSCALD} arise from  updating the BFs
        in \eqref{upadte-v}, which requires computing 
        $(\Ab_n^{k_\ell} + \lambda \Ib)^{-1}$ of $\mathcal{O}(N_t^3)$ complexity. In practice, this 
        could become prohibitive for a large number of transmit antennas.
        
        Interestingly, observing that {\small $\Ab_n^{k_\ell} = \widetilde{\Ab}_n + \mu_{k_\ell} \Cb_n^{k_\ell}$} with {\small $ \widetilde{\Ab}_n \triangleq \sum_{j_\ell \in \Sc_n^\ell}
        (\zeta_n^{j_\ell})^2
        \hat{\hb}_n^{\ell,j_\ell}
        (\hat{\hb}_n^{\ell,j_\ell})^{\rm H}
        +
        ( \sum_{j_\ell \in \Sc_n^\ell} (\zeta_n^{j_\ell})^2 )
        \Db^\ell_n$} that is unchanged during the dual updates and {\small$\Cb_n^{k_\ell}$} being a rank-one matrix, the computation of {\small$(\Ab_n^{k_\ell} + \lambda \Ib)^{-1}$} can be greatly simplified by exploiting the Woodbury matrix identity, i.e.,
        \begin{small}
        \begin{equation} \label{eq:inv_computS}
        \begin{aligned}
        &(\widetilde{\Ab}_n + \lambda \Ib + \mu_{k_\ell} \Cb_n^{k_\ell})^{-1}
        =
        (\widetilde{\Ab}_n + \lambda \Ib)^{-1}  \\
        &\quad
        -
        \frac{
        \mu_{k_\ell}
        (\widetilde{\Ab}_n + \lambda \Ib)^{-1}
        \ab(\hat{\theta}_{n}^{\ell,k_\ell})
        \ab^{\rm H}(\hat{\theta}_{n}^{\ell,k_\ell})
        (\widetilde{\Ab}_n + \lambda \Ib)^{-1}
        }{
        1 + \mu_{k_\ell}
        \ab^{\rm H}(\hat{\theta}_{n}^{\ell,k_\ell})
        (\widetilde{\Ab}_n + \lambda \Ib)^{-1}
        \ab(\hat{\theta}_{n}^{\ell,k_\ell})
        }.
        \end{aligned}
        \end{equation}
        \end{small}

        \vspace{-0.3cm}
        \begin{Rmk} \label{Rmk1}
        In each dual iteration, 
        $(\widetilde{\Ab}_n + \lambda^{(r,s)} \Ib)^{-1}$
        is computed only once and applied to all scheduled users. 
        Therefore, \eqref{eq:inv_computS} reduces the per-user computational complexity 
        of the BF update from 
        $\mathcal{O}(N_t^3)$ in \eqref{upadte-v} to $\mathcal{O}(N_t^2)$, 
        significantly improving the scalability of the proposed algorithm.
        \end{Rmk}

        }

\vspace{-0.4cm}
\section{CKM construction by LSCM} \label{sec:CKM_cons}
		In this section, we briefly introduce the CKM construction.
		First, each BS collects RSRP measurements tagged with user locations, which can be obtained through offline campaigns such as dedicated drive tests, and continuously updated with fresh online user feedback to adapt to environmental dynamics\footnote{The BS does not need to collect measurements from all possible user locations, which is not practical. Instead, exploiting spatial consistency of local wireless channels \cite{CKM_tut24_Zengyong}, the estimates of the channel parameters for locations without measurements can be inferred by interpolation or extrapolation.}. 
		Using RSRP measurements, propagation path characteristics such as the APS are extracted to construct the CSI estimate. The BS stores the CSI estimates with the corresponding user locations, thereby establishing a CKM that provides location-to-CSI mapping.

		Specifically, LSCM techniques \cite{zhang2023physics} are leveraged to extract the APS statistics from RSRP data. Because this process is identical across all BSs, the indices of BSs and users $\ell, j_m$ will be omitted hereafter. As illustrated in Fig. \ref{fig:CKM}, the coverage area $\Ac$ of a BS is divided into grids $i_g = 1, \dots, N_g$, each corresponding to a potential user location. The DL angular spread $[\theta_{min}, \theta_{max}]$ is discretized uniformly into $N_{\theta}$ bins: $\Theta \triangleq \{\theta_1, \dots, \theta_{i_a}, \dots, \theta_{N_{\theta}}\}$ with $\theta_{i_a} = \theta_{min} + (i_a-0.5)\frac{\theta_{max} - \theta_{min}}{N_\theta}, i_a = 1, \dots, N_{\theta}$. With the transmit steering vector $\ab(\theta_{i_a}) \in \!\Cbb^{N_t}$ defined in \eqref{a_theta}, the DL channel from the BS to a user at the $i_g$-th grid is modeled as
		
        \vspace{-0.15cm}
		\begin{small}\begin{equation}\label{eq:DL_ch}
				\hb_{i_g} (t) = \sum\nolimits_{i_a = 1}^{N_{\theta}} \alpha_{i_g, i_a}(t) \ab(\theta_{i_a}), 
			\end{equation}
		\end{small}
		
		\vspace{-0.15cm}
		{\noindent}where $\alpha_{i_g, i_a}(t)$ is the complex gain of the path corresponding to the $i_a$-th angular bin at time $t$. 
		
		\begin{figure}[t] 
			\centering	
			{\includegraphics[width=0.48\textwidth]{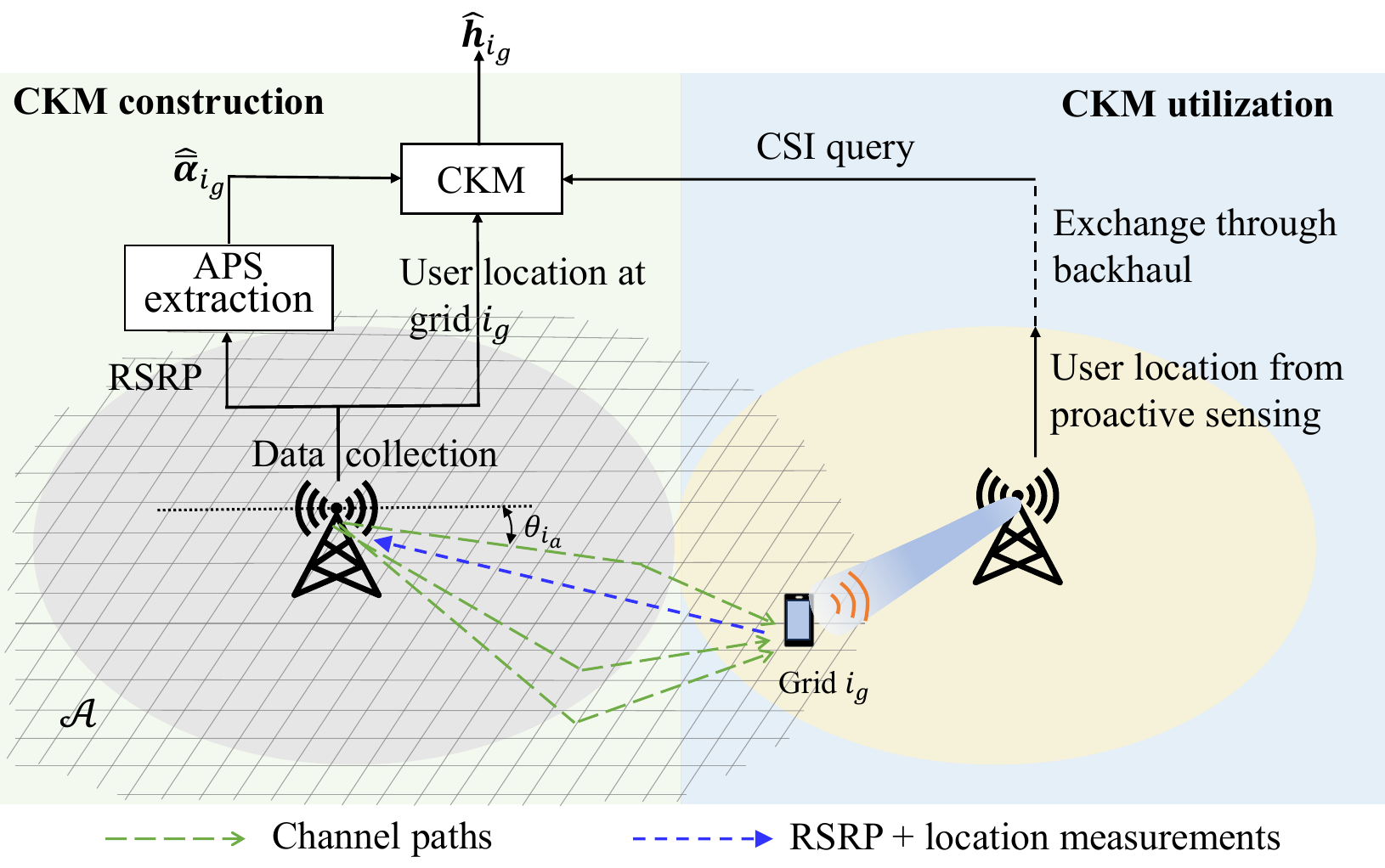}}\vspace{-0.15cm}	
			\caption{Illustration of the CKM construction at the left BS, whose coverage area is divided into grids. The shadowed circle represents the area where users are typically associated with the BS centered at the area. In the CKM utilization, the left BS queries its CKM with the user locations exchanged from the neighboring (right) BS to obtain the inter-cell CSI estimates.} 
			\label{fig:CKM} \vspace{-0.2cm}
		\end{figure}
		
		Assume that the APS is stationary over $T$ successive epochs, the expected APS of the $i_a$-th angular bin at grid $i_g$ over these epochs is
        $
				\bar \alpha_{i_g, i_a} = \Ebb(|\alpha_{i_g, i_a}(t)|^2), i_g = 1, \dots, N_g.
		$
		Stacking these yields the expected channel APS vector
		$
				\bar \alphab_{i_g} = [\bar\alpha_{{i_g}, 1}, \dots, \bar\alpha_{{i_g}, N_\theta}]^\top \in \Rbb^{N_\theta}.
		$
		
		In practice, each BS regularly sends directional reference signal beams to the user for the RSRP measurement. Let the predefined transmit precoding matrix, including $N_b$ beams, be
		$
				\Wb \triangleq \frac{1}{\sqrt{N_t}}[\wb_1, \dots, \wb_{N_b}] \in \Cbb^{N_t \times N_b}, 
		$
		where $\wb_{b} \in \Cbb^{N_t}$ is the codeword associated with the $b$-th beam. For the $b$-th beam, the RSRP measured at grid $i_g$ is $r_{i_g, b}(t) = P_T |\wb_b^H \hb_{i_g}(t)|^2/\sqrt{N_t}$. Collectively, the overall RSRP measurements are 
		
        \vspace{-0.4cm}
		{\small \begin{align}\label{eq:v_rsrp}
				\!\!\!\rb_{i_g}(t) & = [r_{i_g, 1}, \dots, r_{i_g, N_b}]^{\top}  =  P_T |\Wb^H \hb_{i_g}(t)|^2 \in \Rbb^{N_b},
		\end{align}}

        \vspace{-0.1cm}
		{\noindent}where $|\cdot|$ is element-wise modulus.

		Due to the limited scattering effect of mmWave propagation, most of the energy from the multi-path signals is concentrated within a limited number of angular bins. Accordingly, the number of effective paths $N_p \ll N_\theta$, and $\bar\alphab_{i_g}$ is a sparse vector. Therefore, given the measurements $\bar \rb_{i_g}$ fed back from the user, the BS can extract the expected APS $\bar\alphab_{i_g}$ by solving the following sparse recovery problem \cite{zhang2023physics}
		\vspace{-0.35cm}
		
		\begin{small}\begin{subequations} \label{p:APS_extract}
				\begin{align}
					\hat{\bar\alphab}_{i_g} = \arg\min\nolimits_{\bar\alphab_{i_g} \succeq \mathbf{0}} & \ \| \bar \rb_{i_g} - P_T (|\Ab^H \Wb|^2)^T \bar \alphab_{i_g}\|^2, \\
					\text{s.t.} & \ \| \bar \alphab_{i_g}\|_0 \le N_p,
				\end{align}
			\end{subequations}
		\end{small}
		
		\vspace{-0.2cm}
		{\noindent}where $\Ab = [\ab(\theta_1), \dots, \ab(\theta_{i_a}), \dots,  \ab(\theta_{N_\theta})]\in \Cbb^{N_t \times N_{\theta}}$. With $\hat{\bar\alphab}_{i_g}$, the channel estimate for grid $i_g$ is reconstructed as\footnote{Compared with \eqref{eq:DL_ch}, $\hat \hb_{i_g}$ recovers the amplitude of the expected complex gain of the paths, though the phase information is lost due to its absence in the measured RSRPs \eqref{eq:v_rsrp}. However, the estimate constructed based on the APS statistics still captures the large-scale propagation feature under the limitation of practical cellular channel measurement, bringing a distinct performance to multi-cell coordinated transmission, as shown in Sec. \ref{sec:simu}. } 
		\vspace{-0.2cm}
		
		\begin{small}\begin{equation}\label{eq:h_est}
				\hat \hb_{i_g} = \sum\nolimits_{i_a = 1}^{N_{\theta}} \sqrt{\hat{\bar\alpha}_{i_g, i_a}} \ab(\theta_{i_a}).  
			\end{equation} 	
		\end{small}
		
		\vspace{-0.1cm}
		{\noindent}Following this procedure, a CKM including the channel estimates $\{\hat \hb_{i_g}\}_{i_g = 1, \dots, N_g}$ for BS $\ell \in \Lc$ can be established.

		\vspace{-0.3cm}
		\section{Simulation Results \label{sec:simu}}
		
		In this section, extensive simulations 
        %{based on synthetic data and DeepMIMO\footnote{ https://www.deepmimo.net/}} dataset 
        are conducted to validate our proposed SD-USCB framework. The network includes $L = 3$ cells, each with a BS equipped with $N_t=32$ transmit antennas and $N_r=64$ receiving antennas. The BSs are located at coordinates $(0, 0)$ $(180, 0)$, and $(90, 90\sqrt{3})$ (unit: $\text{m}$). There are $|\Uc_\ell| = 40$ users associated with each BS, {which schedules a maximum of 18 users in each epoch, unless stated otherwise.} The velocity of each user is set to $v = 20 \text{m/s}$. The transmit power budget of each BS is $P_\ell = 36\,\text{dBm}$.  The carrier frequency of the ISAC signal is $f_c = 30 \, \text{GHz}$.  The length of each epoch and the backhaul delay are set to $\Delta T = 20 \, \text{ms}$ and $T_d = 4 \, \text{ms}$, respectively. The CKM grid size is chosen to match the maximum user displacement per epoch, $v \cdot \Delta T$, maintaining CSI fidelity while avoiding high CKM computational cost; in our simulations, this corresponds to $0.4\text{m}\times 0.4\text{m}$. The DL communication intra-cell channel in the $n$-th epoch is generated as
		\vspace{-0.3cm}
		
		\begin{small}\begin{align}           
				\boldsymbol{h}^{\ell,k_\ell}_n(t)=&~\sqrt{{\alpha}^{\ell,k_\ell}_{1,n}(t)} e^{jw^{\ell,k_\ell}_{1,n}}\ab\left(\phi^{\ell,k_\ell}_{1,n}\right) \notag \\
				&+ \sum\nolimits_{q=2}^{N_p }\sqrt{{\alpha}^{\ell,k_\ell}_{q,n}(t)}e^{jw^{\ell,k_\ell}_{q,n}}\ab\left(\phi^{\ell,k_\ell}_{q,n}\right),
			\end{align}
		\end{small}
		
		{\noindent}including one LoS path and $N_p-1$ NLoS paths.
		The LoS path-loss is modeled as ${\alpha}^{\ell,k_\ell}_{1,n}(t) = \alpha_0 (d^{\ell,k_\ell}_n)^{-\eta}$, where $\alpha_0 = -60 \text{dB}$ is the reference path-loss at the distance of 1m and $\eta = 2$. The AoD angle of the NLoS $\phi^{\ell,k_\ell}_{q,n}$ is randomly generated from $\mathcal{U}(-\pi, \pi)$, and the NLoS path loss satisfies $\alpha^{\ell,k_\ell}_{1,n}/\alpha^{\ell,k_\ell}_{q,n} \in \mathcal{U}(4, 9), q > 1$, with $\mathcal{U}(a, b)$ being the uniform distribution within $[a,b]$. Besides, the inter-cell channel $\hb^{m,k_\ell}_n(t)$ generated by $N_p$ NLoS paths
		\vspace{-0.2cm}
		
		\begin{small}\begin{align}           
				\boldsymbol{h}^{m,k_\ell}_n(t)=
				\sum\nolimits_{q=1}^{N_p} \sqrt{{\alpha}^{m,k_\ell}_{q,n}(t)} e^{jw^{m,k_\ell}_{q,n}}\ab\left(\phi^{m,k_\ell}_{q,n}\right),
			\end{align}
		\end{small}
		
		{\noindent}where the phases $\{w^{m,k_\ell}_{q,n}\}$ are randomly sampled from $\mathcal{N}(0,(\pi/5)^2)$ and $N_p = 4$. The power of the communication noise $\sigma_c^2$ satisfies ${\small\frac{P_\ell \sum_{m\in\mathcal{L}} \sum_{\ell\in\mathcal{L}}\sum_{k_\ell \in \mathcal{U}_\ell}||\hb^{m,k_\ell}_1||^2}{\sum_{\ell \in \mathcal{L}}L|\mathcal{U}_\ell|N_t\sigma_c^2} = 15 \text{dB}}$. 
        For the CKM construction, the precoding matrix $\Wb$ for RSRP measurement is generated from discrete Fourier transform (DFT) matrices with $N_b = 64$, and the measurement noise power equals $\sigma_c^2$. With RSRP measurements, the OLS algorithm \cite{lu2022recovery} is applied to solve problem \eqref{p:APS_extract} to recover the corresponding expected APS values. 
        For the sensing, the parameters appeared in \eqref{eq:SINRs} are set as $a_\theta = 0.1$, $a_\tau = 6.7 \times 10^{-7}$, $a_\mu = 0.1$, $G = 10$, and $\bar c = 1$.
		
        \subsection{Performance of Proposed DualOpt-Based BF Algorithm \ref{alg:BF_FPSCALD}}

        \begin{figure}[t] 
			\centering	{\includegraphics[width=0.4\textwidth]{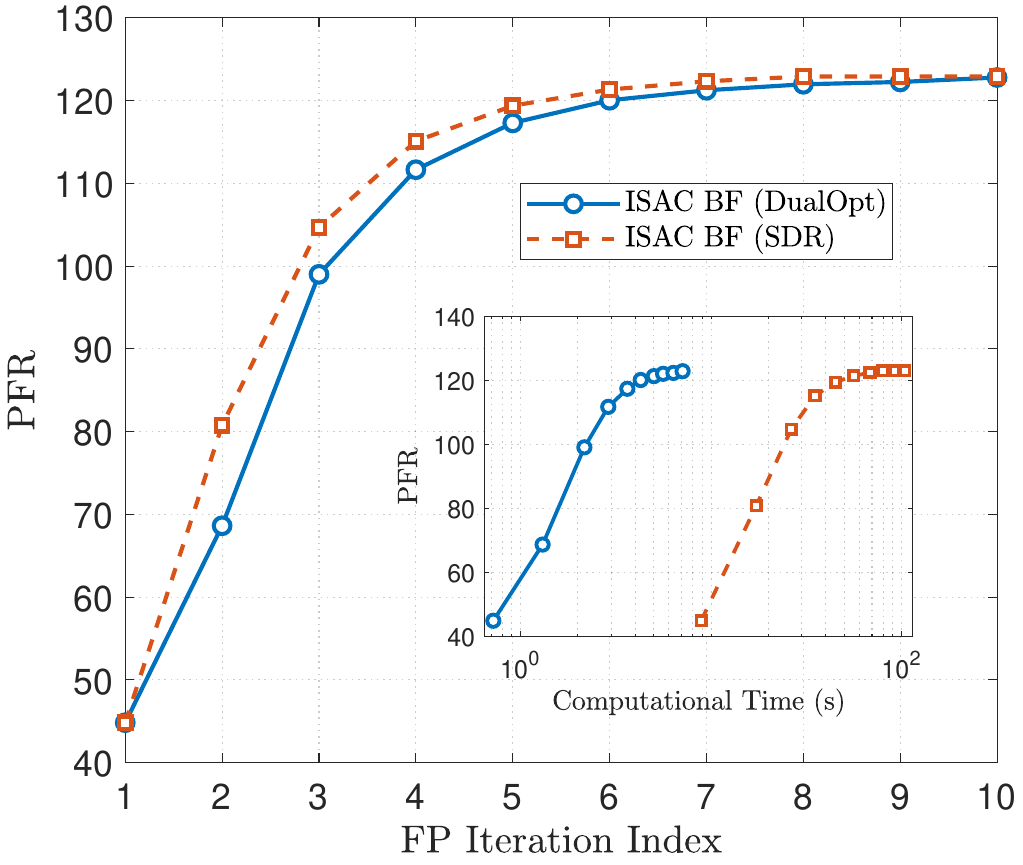}} \vspace{-0.25cm}
			\caption{PFR achieved by different BF algorithms versus FP iteration.} 
			\label{fig:time}
		\end{figure} 

        Fig. \ref{fig:time} shows the convergence behavior of our proposed DualOpt-based ISAC BF Algorithm \ref{alg:BF_FPSCALD}, benchmarked against the SDR-based algorithm \cite{SD_USBF_icassp24}. One can observe that both algorithms converge within about 10 iterations. Upon convergence, the achieved PFR of our proposed Algorithm \ref{alg:BF_FPSCALD} is almost the same as that of the SDR-based approach. However, as evidenced in Fig. \ref{fig:time}, Algorithm \ref{alg:BF_FPSCALD} substantially reduces computational time. This owes to that our design of Algorithm \ref{alg:BF_FPSCALD} directly tackles the problem \eqref{p:BF_ISAC_opt1} over vector variables, thereby circumventing relaxation to a high-dimension SDP. Moreover, Algorithm \ref{alg:BF_FPSCALD} fully exploits the problem structure as highlighted in Remark \ref{Rmk1} and takes only first-order computations, making it very efficient. In Table \ref{Th:1}, the averaged computational time over 10 epochs of the two algorithms is compared under different numbers of users $|\Sc_n| = \sum_{\ell} |\Sc_n^\ell|$ scheduled in the overall network. Compared with its counterpart, Algorithm \ref{alg:BF_FPSCALD} achieves more than 7-fold reduction in computational time for $|\Sc_n| = 18$, with even greater gains observed as $|\Sc_n|$ increases. These results confirm that our proposed ISAC BF Algorithm \ref{alg:BF_FPSCALD} not only enhances PFR effectively but also delivers significant computational efficiency.

        \begin{table}[t]
		\centering
		\caption{Average Computational Time (s) per BS per epoch}
		\vspace{-0.2cm}
		\begin{tabular}{l|llll}\hline
			Algorithm & $|\Sc_n|= 18$ & $|\Sc_n|=36$ & $|\Sc_n| = 54$  \\ \hline
			ISAC BF (SDR)   & 6.7 & 19  & 33         \\
			ISAC BF (DualOpt)    & 0.9   & 1.1    & 2.4           \\ \hline
		\end{tabular}
		\label{computation1} \vspace{-0.1cm}
	    \end{table}
		\vspace{-0.1cm}
		\subsection{Effectiveness of {Proposed SD-USCB} Framework}
		\vspace{-0.05cm}
		\begin{figure}[t] 
			\centering	{\includegraphics[width=0.4\textwidth]{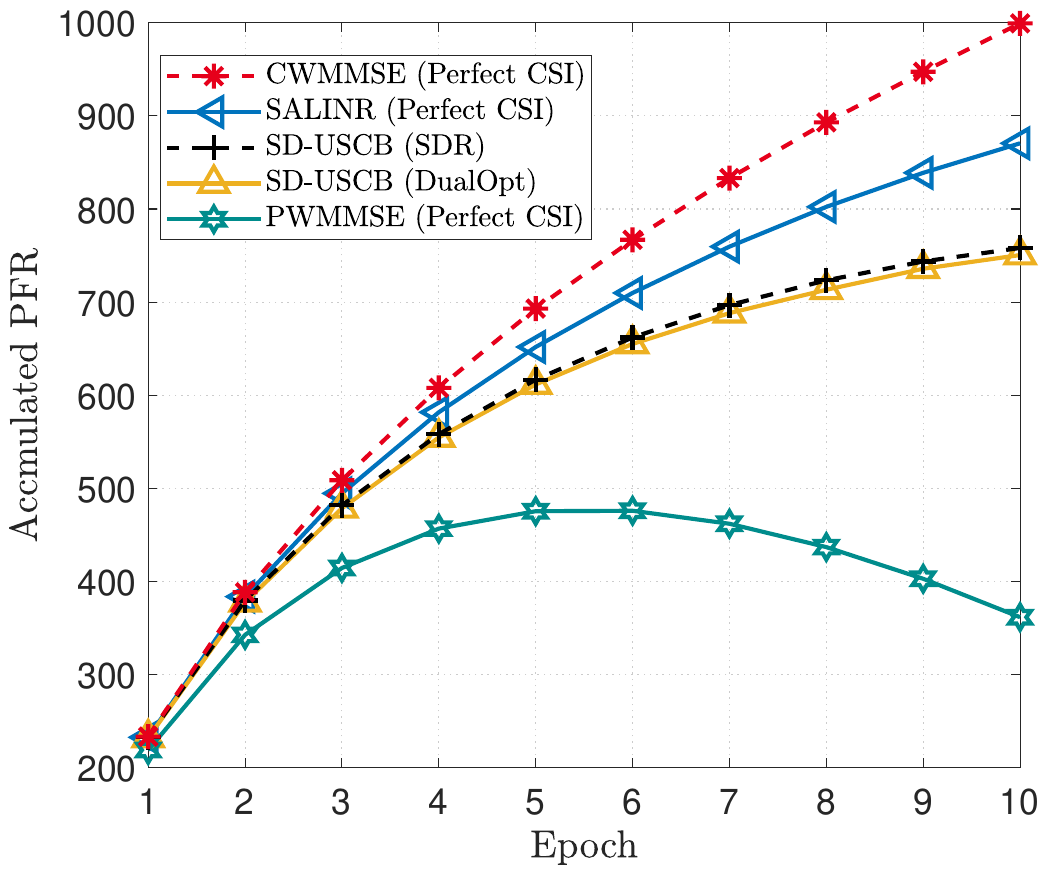}} \vspace{-0.25cm}
			\caption{Accumulated PFR achieved by different schemes versus epoch.} 
			\label{fig:Proposed without aging}
		\end{figure}

		In Fig. \ref{fig:Proposed without aging}, we show the accumulated PFR achieved by our proposed SD-USCB framework. For comparison, we also include three schemes for BF design: the classic centralized weighted minimum mean square error (CWMMSE) algorithm \cite{christensen2008weighted}, the per-cell WMMSE (PWMMSE) algorithm, and the FP algorithm that solves the SALINR-based virtual PFR distributively. All benchmark schemes are implemented under perfect CSI, and their BF optimization does not account for the sensing requirements. Besides, they all employ the scheduling algorithm in \cite{yoo2006optimality} for user scheduling. 
		
		As anticipated, CWMMSE as an ideal scheme achieves the highest accumulated PFR, since it has perfect intra-cell and inter-cell CSI. In contrast, PWMMSE yields the lowest PFR, as it only alleviates the intra-cell interference while completely ignoring inter-cell interference. This also renders extremely low rates for certain users, which after logarithmic transformation in \eqref{eq:obj_PFR_SINR0}, produce negative values -- explaining why PWMMSE’s accumulated PFR declines after epoch 6. Meanwhile, SALINR (Perfect CSI) approaches CWMMSE in performance. Notably, our SD-USCB framework delivers accumulated PFR nearly on par with SALINR and substantially surpasses PWMMSE, demonstrating its efficacy in improving the network performance under practical limitations. 
        
        Critically, in contrast to CWMMSE and SALINR, SD-USCB significantly reduces CSI acquisition overhead and cross-cell information exchange by integrating proactive sensing with CKM -- a crucial advantage for practical networks with bandwidth-limited backhauls. Moreover, SD-USCB’s distributed processing markedly alleviates the computational burden at a single node (CU) required by CWMMSE, offering superior scalability in large-scale networks. 
        
        %Notably, within the SD-USCB framework, the performance of the DualOpt approach closely matches that of the SDR-based method, demonstrating the effectiveness and efficiency of the dual-opt solution.
		\vspace{-0.25cm}
		\subsection{Impact of Sensing Error Threshold}
		\begin{figure}[t] 
			\centering	{\includegraphics[width=0.4\textwidth]{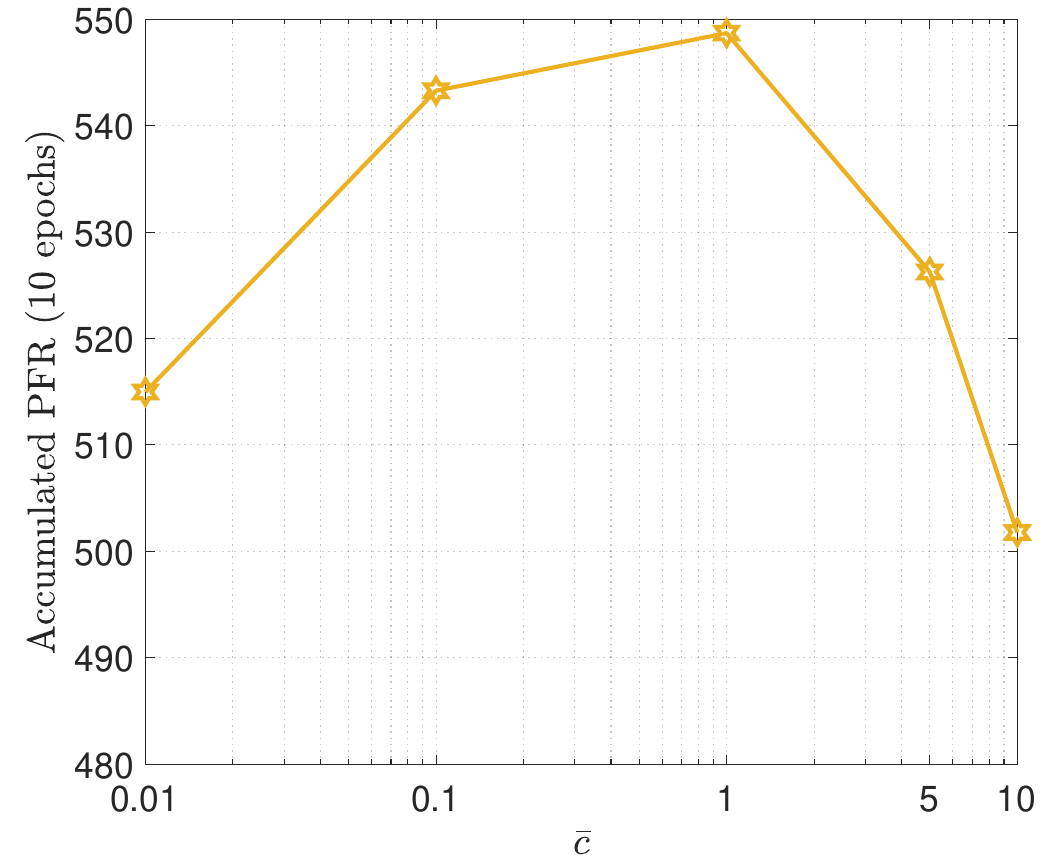}} \vspace{-0.25cm}
			\caption{Accumulated PFR achieved by SD-USCB under different sensing thresholds.} 
			\label{fig: Proposed BF c}
		\end{figure}
		In Fig. \ref{fig: Proposed BF c}, the impact of different values of the sensing error threshold $\bar{c}$ in \eqref{p:BF_ISAC_opt} on the PFR achieved by our proposed SD-USCB is examined. To isolate the influences from other factors, we apply Algorithm \ref{alg:PFZFG} to schedule $  |\Sc_n^\ell| = 18$ users in the 1st epoch, maintain this scheduling set throughout the subsequent 9 epochs, and observe the accumulated FPR achieved in epoch 10. Interestingly, one can see that a stringent sensing requirement $\bar{c} =0.01$ would result in a decline in PFR performance. This is because BSs in the network need to allocate more resources to meet the sensing requirements, while fewer resources are left for communication. Conversely, when the threshold of sensing error is too large, i.e., $\bar{c}=10$, the accuracy of sensed user kinematic parameters is not guaranteed. As a result, the estimated user locations would have large errors. When these locations are input to the CKMs for CSI query, the accuracy of the attained CSI estimates will degrade, leading to a decreased PFR. In comparison, when the threshold $\bar{c}$ is set between 0.1 and 1, the performance is better, implying this range ensures sufficient sensing accuracy while leaving more resources available for communication.

		\begin{figure}[t] 
			\centering	{\includegraphics[width=0.4\textwidth]{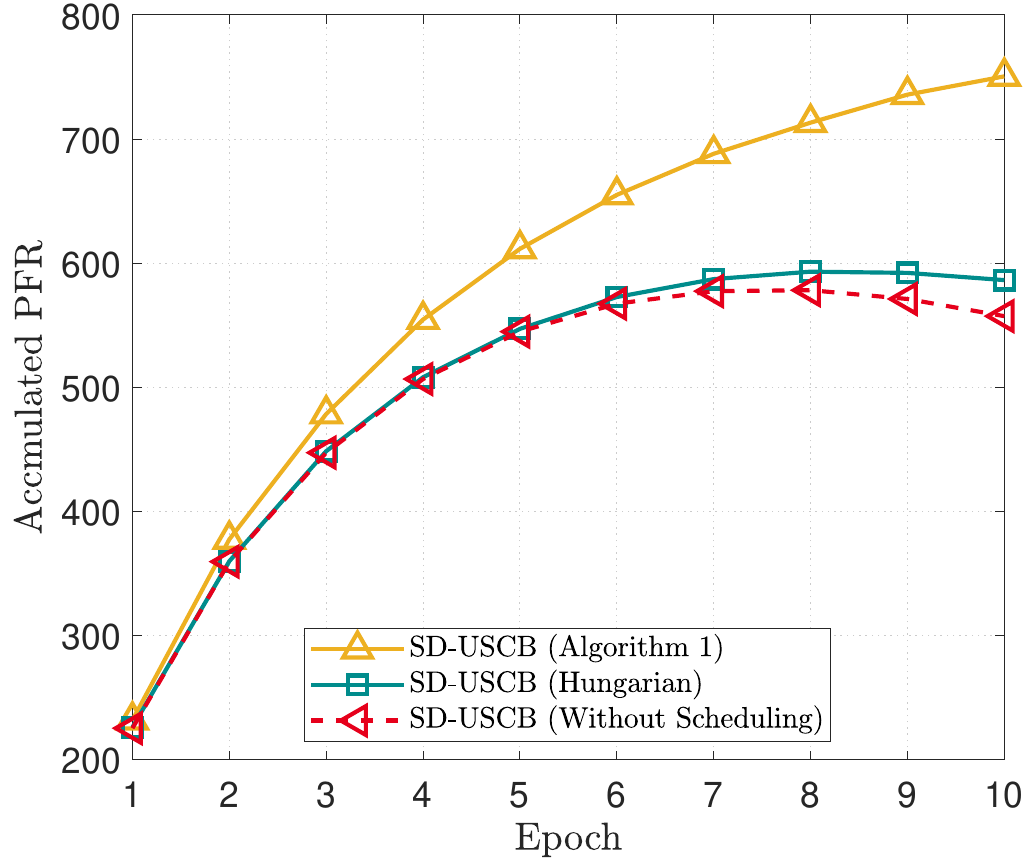}} \vspace{-0.25cm}
			\caption{Accumulated PFR of different scheduling algorithms versus epoch.} 
			\label{fig:scheduling}
		\end{figure}
		\vspace{-0.25cm}
		\begin{figure}[t] 
			\centering	{\includegraphics[width=0.4\textwidth]{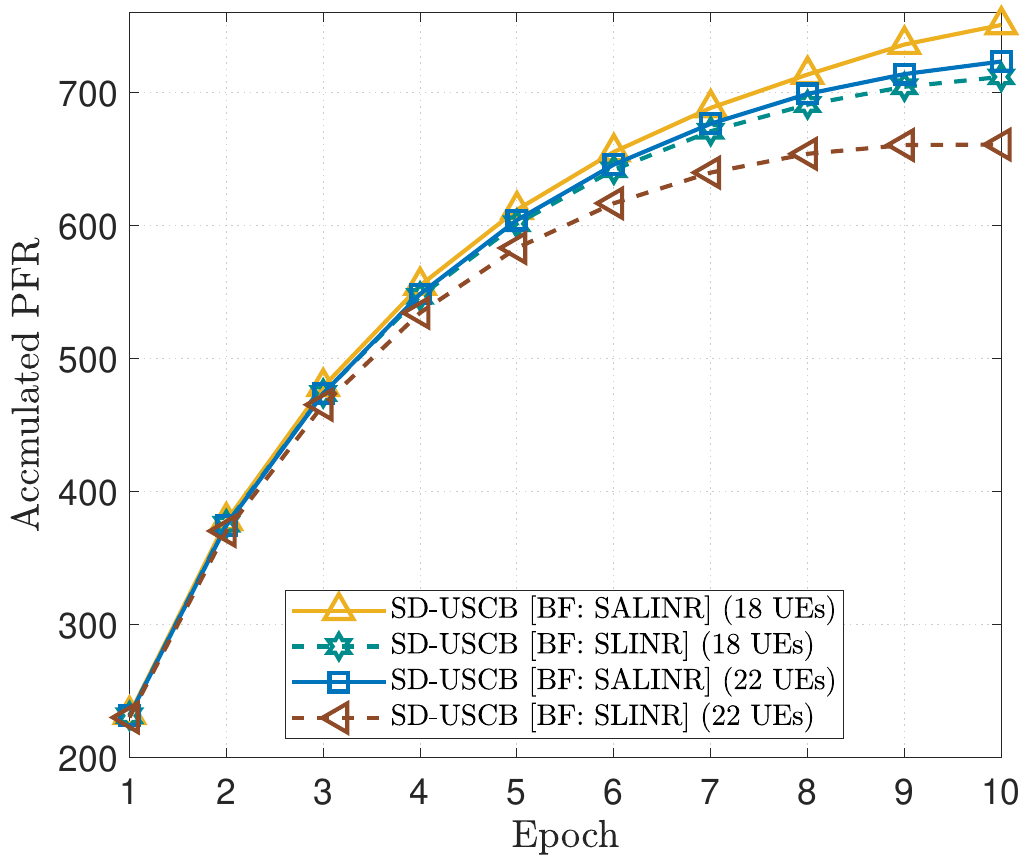}} \vspace{-0.25cm}
			\caption{Accumulated PFR vs. epoch: SALINR vs. SLINR BF.} 
			\label{fig:SALINR}
		\end{figure}

		\vspace{-0.05cm}
		\subsection{Impact of Scheduling}
        To examine the efficacy of our developed scheduling Algorithm \ref{alg:PFZFG} in the SD-USCB framework, we compare SD-USCB against two benchmark schemes that modify only the user scheduling algorithm while keeping the rest of the framework unchanged.  Specifically, `SD-USCB (No scheduling)' selects all users in $\Uc_\ell, \ell \in \Lc$ for transmission in each epoch, while `SD-USCB (Hungarian)' replaces Algorithm \ref{alg:PFZFG} by the Hungarian algorithm \cite{kuhn1955hungarian}.
		
		For SD-USCB (without scheduling), since all users are scheduled in each epoch, the ISAC beams pointing to these geographically nearby users overlap easily, causing severe interference. As a result, BSs need to invest more resources to meet the sensing requirements. Meanwhile, the interference deteriorates the communication performance, so its achieved PFR is the lowest.
		While the accumulated PFR of SD-USCB (Hungarian) %which applies the Hungarian scheduling algorithm \cite{kuhn1955hungarian} in our framework, %determines the set of scheduled users by alternately optimizing virtual beamformers and the scheduling process until convergence. However, this alternating optimization method incurs high computational costs. While 
        is higher than that of SD-USCB (Without scheduling), it remains significantly lower than SD-USCB (Algorithm \ref{alg:PFZFG}). One possible reason is that the user set selected by the Hungarian algorithm exhibits stronger user channel correlations than those in Algorithm \ref{alg:PFZFG}, which leads the BS to cost more resources for interference alleviation. These results validate the efficacy of the scheduling Algorithm \ref{alg:PFZFG}.

        \vspace{-0.3cm}
        \subsection{SALINR vs SLINR in BF} \label{sec:simu_SALINR}
		% {\bf SALINR vs SLINR in BF}:  
        Fig. \ref{fig:SALINR} compares the performance of our proposed SD-USCB framework with BF based on SALINR and that with BF based on conventional SLINR \cite{li2022decentralized}. As pointed out in Sec. \ref{sec:SLINR}, SLINR approximates the original incoming cross-cell interference to a serving user by the leakage from that user to scheduled users in other cells. In contrast, SALINR approximates this interference using the arithmetic mean of the interference power generated by BS $\ell$ to all scheduled users in other cells.  While the arithmetic mean operation is straightforward, the BF based on SALINR delivers a remarkable performance improvement. As shown in Fig. \ref{fig:SALINR}, under the same number of scheduled users per cell $|\Sc_n^\ell|$, SD-USCB [BF: SALINR] outperforms SD-USCB [BF: SLINR], with the performance gap widening over successive epochs. Moreover, with $|\Sc_n^\ell|$ increasing from 18 to 22, the performance gain of SD-USCB [BF: SALINR] is more pronounced. This is because SALINR can provide a more accurate approximation of the original SINR, as analytically established in Theorem \ref{Th:1}. The underlying insight is that SALINR  reduces the discrepancy between the leakage from a single user and the interference that user experiences by considering the average interference leakage at the cell level. While Theorem \ref{Th:1} is derived under Rayleigh channels, the results in Fig. \ref{fig:SALINR} demonstrate that SALINR-based BF also achieves better performance in general multi-path channels, confirming its superiority and resilience for D-MCRA design.

		\vspace{-0.25cm}
		\section{Conclusion \label{sec:conclusion}}
		
		In this paper, we have investigated the D-USCB design in multi-cell mmWave networks. To overcome the challenges of inter-cell CSI acquisition and the large information exchange overhead through the limited-bandwidth backhauls, we have proposed a new SD-USCB framework. Our framework includes: (i) a novel combination of the CKM and ISAC techniques for acquiring the inter-cell CSI with low backhaul overhead; (ii) a leakage-based metric SALINR to achieve fully distributed BF optimization, with its superiority in SINR approximation being theoretically established; (iii) a heuristic distributed user scheduling algorithm to mitigate the intra-cell interference; and (iv) an efficient dualOpt-based ISAC BF algorithm that improves the network performance while greatly reducing the computational time. Besides, we presented an approach to construct CKMs from RSRP measurements via LSCM techniques. By extensive simulations, we have demonstrated that our proposed framework can effectively enhance the network throughput with greatly reduced overhead for CSI acquisition and information exchange.	
		\appendices
		\vspace{-0.3cm}
		
		\section{Proof of Theorem \ref{Th:1}} \label{App:proof}
			
			Given a set of fixed beamformers $\{\vb_{n}^{k_\ell}\}$ with $\|\vb_n^{k_\ell}\|^2 = P, k_\ell \in \Uc_\ell, \ell \in \Lc$. 
			For independent Rayleigh fading channels, 
			\(\hb_n^{m,k_\ell} \sim \mathcal{CN}(\mathbf{0}, \Ib_{N_t})\), and the inner product of $\hb_n^{m,k_\ell}$ and $\vb_n^{m,k_\ell}$
			will satisfy
			\vspace{-0.2cm}
			
			\begin{small}
				\begin{equation}
					(\hb_n^{m,k_\ell})^{\rm{H}} \vb_n^{t_m} \sim \mathcal{CN}(0, P), \vspace{-0.3cm}
				\end{equation}
			\end{small}
			
			\noindent Moreover, the square of its magnitude follows an exponential distribution $|(\hb_n^{m,k_\ell})^{\rm{H}} \vb_n^{t_m}|^2 \sim \mathrm{Exp}(P).$
			
			As $\Ic_n^{k_\ell}$ is the sum of  $M = \sum_{m\neq \ell} \sum_{t_m} q_n^{t_m}$ independent 
			\(\mathrm{Exp}(P)\) random variables, it follows a Gamma distribution as
			\vspace{-0.5cm}
			
			\begin{small}
				\begin{equation} \label{eq:ICI}
					\Ic_n^{k_\ell} \sim \mathrm{Gamma}(M, P), \vspace{-0.2cm}
				\end{equation}
			\end{small}
			
			\noindent whose mean is \(\E[\Ic_n^{k_\ell}] = M P\) and variance is
			\(\mathrm{Var}[\Ic_n^{k_\ell}] = M P^2\).
			Similarly, the conventional leakage term $\hat{L}_n^{k_\ell}$ will satisfy 
			
			\vspace{-0.3cm}
			\begin{small} 
				\begin{equation} \label{eq:Leak_conv}
					\hat{L}_n^{k_\ell} \sim \mathrm{Gamma}(M, P), \vspace{-0.2cm}
				\end{equation}
			\end{small}
			
			\noindent while our proposed averaged leakage $\tilde{L}_n^{k_\ell}$ satisfies		
			\vspace{-0.2cm}
			
			\begin{small}
				\begin{equation} \label{eq:Leag_avg}
					\tilde{L}_n^{k_\ell} \sim \mathrm{Gamma}(M, P/\sqrt{|\Sc_n^\ell|}), \vspace{-0.2cm}
				\end{equation}
			\end{small}

            \noindent with $|\Sc_n^\ell| = \sum\nolimits_{j_\ell\in\Uc_\ell}{q^{j_\ell}_n}$.
            
			For $ R_n^{k_\ell}(\Ic_n^{k_\ell}) = \log\left(1 + {S_n^{k_\ell}}/
			({T_n^{k_\ell} + \Ic_n^{k_\ell} + \sigma_c^2})\right)$, it is not difficult to verify that its derivative ${R_n^{k_\ell}}'(\Ic_n^{k_\ell})$ is non-increasing over $\Ic_n^{k_\ell} \ge 0$. Therefore, the Lipschitz constant of $R_n^{k_\ell}(\Ic_n^{k_\ell})$ can be bounded as
			\vspace{-0.2cm}
			
			\begin{small}
				\begin{subequations}
				    \begin{align}\label{zeta}
					\sup\nolimits_{\Ic_n^{k_\ell} \ge 0}|{R_n^{k_\ell}}'(\Ic_n^{k_\ell})| &\le \frac{S_n^{k_\ell}}{(\sigma_c^2+T_n^{k_\ell}) (\sigma_c^2 +T_n^{k_\ell}+S_n^{k_\ell})}  \\
                    &\le \frac{S_n^{k_\ell}}{\sigma_c^2 (\sigma_c^2 +S_n^{k_\ell})} \triangleq Z_n^{k_\ell},
				\end{align} 
				\end{subequations}
			\end{small}\vspace{-0.2cm}
			
			\noindent with $S_n^{k_\ell} = |(\hb^{\ell,k_\ell}_{n} )^{\rm{H}} \vb_{n}^{k_\ell}|^2\sim \mathrm{Exp}(P)$ and $T_n^{k_\ell}\ge 0$.
			Define the rate approximation errors as $
			\zeta_{n,1}^{k_\ell} = \hat{R}_n^{k_\ell} - R_n^{k_\ell}$ and $\zeta_{n,2}^{k_\ell} = \tilde{R}_n^{k_\ell} - R_n^{k_\ell}$,  and the ICI approximation errors as $\epsilon_{n, 1}^{k_\ell} = \hat{L}_n^{k_\ell} - \Ic_n^{k_\ell}$ and $\epsilon_{n, 2}^{k_\ell} = \tilde{L}_n^{k_\ell} - \Ic_n^{k_\ell}$. Based on the Lipschitz continuity of $R_n^{k_\ell}(\Ic_n^{k_\ell})$, we have 
			\vspace{-0.2cm}
			
			\begin{small}
				\begin{equation}\label{eq:lipschitz_single}
					|\zeta_{n, i}^{k_\ell}| \le Z_n^{k_\ell} |\epsilon_{n,i}^{k_\ell}|, i = 1, 2. \vspace{-0.2cm} 
				\end{equation}
			\end{small}
			
			\noindent Further, applying the Cauchy–Schwarz inequality yields the expected error bound of rate approximations
			\vspace{-0.2cm}
			
			\begin{small}
				\begin{equation}\label{eq:lipschitz_sum}
					\E_{\{\hb_n^{k_\ell}\}}\big[\big| \zeta_{n,i}^{k_\ell}\big|\big]
					\! \le \!
				 (\bar Z_n^{k_\ell})^{1/2} \big(  \E_{\{\hb_n^{k_\ell}\}}  [(\epsilon_{n, i}^{k_\ell})^{2}]\big)^{1/2}, i = 1, 2, \!\!
				\end{equation}
			\end{small}
			\vspace{-0.2cm}
			
			\noindent where $\bar Z_n^{k_\ell} =\E_{\{\hb_n^{k_\ell}\}} [ (Z_n^{k_\ell})^2] = \int_{0}^{\infty} \frac{s^2}{ \sigma_{c}^4(\sigma_{c}^2 + s^2)^2}\frac{1}{P}e^{-s/P} ds$. As the true ICI $\Ic_n^{k_\ell}$ and the conventional leakage $\hat{L}_n^{k_\ell}$ follow identical Gamma distributions, with the independence assumption, the expected MSE of the SLINR-based ICI approximation simplifies to the sum of their variances as
			
			\vspace{-0.3cm}
			\begin{small}
				\begin{equation} \label{eq:MSE_leakageConv}
					\E_{\{\hb_n^{k_\ell}\}}[(\epsilon_{n, 1}^{k_\ell})^{2}] = \mathrm{Var}(\hat{L}_n^{k_\ell}) + \mathrm{Var}(\Ic_n^{k_\ell}) = 2MP^2, \vspace{-0.2cm}
				\end{equation}
			\end{small} 
			
			\noindent Similarly, the expected MSE of the SALINR-based ICI approximation will be 
			\vspace{-0.2cm}
			
			\begin{small} 
				\begin{equation} \label{eq:MSE_leakageAvg}
					\!\!\E_{\{\hb_n^{k_\ell}\}}[(\epsilon_{n, 2}^{k_\ell})^{2}]\!=\! \mathrm{Var}(\tilde{L}_n^{k_\ell}) \!+\! \mathrm{Var}(\Ic_n^{k_\ell}) \!=\!  MP^2\!\left(1 +  {1}/{|\Sc_n^\ell|}\right). \vspace{-0.2cm}
				\end{equation}
			\end{small}

			\noindent Apparently, $\E_{\{\hb_n^{k_\ell}\}}[(\epsilon_{n, 2}^{k_\ell})^{2}]  < \E_{\{\hb_n^{k_\ell}\}}[(\epsilon_{n, 1}^{k_\ell})^{2}] $ holds as long as $|\Sc_n^\ell|  >1 $, with the improvement becoming more significant as $|\Sc_n^\ell|$ increases. Substituting \eqref{eq:MSE_leakageConv} and \eqref{eq:MSE_leakageAvg} into \eqref{eq:lipschitz_sum} completes the proof. 
		      \hfill $\blacksquare$
              \vspace{-0.1cm}

		\footnotesize 
		\bibliographystyle{IEEEtran}
		\bibliography{refs_journal_MCRA}

		\ifCLASSOPTIONcaptionsoff
		\newpage
		\fi
		
	\end{document}